\documentclass[sigconf]{acmart}

\usepackage{amsmath,amsfonts}
\usepackage{graphicx}
\usepackage{textcomp}
\usepackage{xcolor}

\usepackage{balance}
\usepackage{bm}
\usepackage{adjustbox}
\usepackage{graphicx,xcolor} 
\usepackage[flushleft]{threeparttable}
\usepackage{balance}
\usepackage{algorithm}
\usepackage{algpseudocode}
\usepackage{multirow}
\usepackage{booktabs}
\usepackage{listings}
\usepackage{xcolor}
\usepackage{amsmath}
\usepackage{algpseudocode}
\usepackage{tabularx,booktabs}
\usepackage{hyperref}
\usepackage{caption}
\usepackage{listings}
\definecolor{codebg}{rgb}{0.96,0.96,0.96}
\definecolor{codebg1}{rgb}{1.0,0.8509803922,0.7490196078}
\definecolor{codebg2}{rgb}{0.7607843137,0.9411764706,0.7843137255}
\definecolor{codered}{rgb}{0.8431372549,0.2274509804,0.2862745098}
\definecolor{codeorange}{rgb}{0.8901960784,0.3843137255,0.03529411765}
\definecolor{codepurple}{rgb}{0.4352941176,0.2588235294,0.7568627451}
\lstdefinestyle{github}{
  language=Python,
  basicstyle=\ttfamily\footnotesize,
  backgroundcolor=\color{codebg},
  keywordstyle=\color{codered}\bfseries,
  deletekeywords=[2]{type},
  emph=[1]{GNN,UnevenDDPIndices,DynamicLoadBalancer},
  emphstyle=[1]{\color{codeorange}\bfseries},
  emph=[2]{True,config,update},
  emphstyle=[2]{\color{codepurple}},
  emph=[3]{get_device,is_uva,get_sample_workers,hybrid_train,train,_train,unified_train},
  emphstyle=[3]{\color{codepurple}\bfseries},
  emph=[4]{w},
  emphstyle=[4]{\color{codebg}},
  captionpos=b,
  breaklines=true,    
}
\newcommand{\reducespace}{\vspace{-2.4mm}} 

\AtBeginDocument{%
  \providecommand\BibTeX{{%
    \normalfont B\kern-0.5em{\scshape i\kern-0.25em b}\kern-0.8em\TeX}}}

\setcopyright{acmcopyright}
\copyrightyear{2024}
\acmYear{2024}
\setcopyright{rightsretained}
\acmConference[CF '24]{21st ACM International Conference on Computing Frontiers}{May 7--9, 2024}{Ischia, Italy}
\acmBooktitle{21st ACM International Conference on Computing Frontiers (CF '24), May 7--9, 2024, Ischia, Italy}\acmDOI{10.1145/3649153.3649191}
\acmISBN{979-8-4007-0597-7/24/05}

\AtBeginDocument{%
  \providecommand\BibTeX{{%
    \normalfont B\kern-0.5em{\scshape i\kern-0.25em b}\kern-0.8em\TeX}}}

\begin{document}
% \fancyhead{}
%%
%% The "title" command has an optional parameter,
%% allowing the author to define a "short title" to be used in page headers.
\title{A Unified CPU-GPU Protocol for GNN Training}

\author{Yi-Chien Lin}
\authornote{Both authors contributed equally to this research.}
\email{yichienl@usc.edu}
\affiliation{%
  \institution{University of Southern California}
  \city{Los Angeles}
  \state{California}
  \country{USA}
}

\author{Gangda Deng}
\email{gangdade@usc.edu}
\authornotemark[1]
\affiliation{%
  \institution{University of Southern California}
  \city{Los Angeles}
  \state{California}
  \country{USA}
}

\author{Viktor Prasanna}
\email{prasanna@usc.edu}
\affiliation{%
  \institution{University of Southern California}
  \city{Los Angeles}
  \state{California}
  \country{USA}
}

% \def\BibTeX{{\rm B\kern-.05em{\sc i\kern-.025em b}\kern-.08em
%     T\kern-.1667em\lower.7ex\hbox{E}\kern-.125emX}}
%% By default, the full list of authors will be used in the page
%% headers. Often, this list is too long, and will overlap
%% other information printed in the page headers. This command allows
%% the author to define a more concise list
%% of authors' names for this purpose.
% \renewcommand{\shortauthors}{Lin et al.}

%%
%% The abstract is a short summary of the work to be presented in the
%% article.
% \begin{abstract}
%   A clear and well-documented \LaTeX\ document is presented as an
%   article formatted for publication by ACM in a conference proceedings
%   or journal publication. Based on the ``acmart'' document class, this
%   article presents and explains many of the common variations, as well
%   as many of the formatting elements an author may use in the
%   preparation of the documentation of their work.
% \end{abstract}

%%
%% The code below is generated by the tool at http://dl.acm.org/ccs.cfm.
%% Please copy and paste the code instead of the example below.
%%
\begin{CCSXML}
<ccs2012>
   <concept>
       <concept_id>10010147.10010169</concept_id>
       <concept_desc>Computing methodologies~Parallel computing methodologies</concept_desc>
       <concept_significance>500</concept_significance>
       </concept>
 </ccs2012>
\end{CCSXML}

\ccsdesc[500]{Computing methodologies~Parallel computing methodologies}

%%
%% Keywords. The author(s) should pick words that accurately describe
%% the work being presented. Separate the keywords with commas.
\keywords{GNN, Unified CPU-GPU protocol, GNN training}

%% A "teaser" image appears between the author and affiliation
%% information and the body of the document, and typically spans the
%% page.

%%
%% This command processes the author and affiliation and title
%% information and builds the first part of the formatted document.
\begin{abstract}
Training a Graph Neural Network (GNN) model on large-scale graphs involves a high volume of data communication and computations.
While state-of-the-art CPUs and GPUs feature high computing power, the Standard GNN training protocol adopted in existing GNN frameworks cannot efficiently utilize the platform resources.
To this end, we propose a novel Unified CPU-GPU protocol that can improve the resource utilization of GNN training on a CPU-GPU platform.
The Unified CPU-GPU protocol instantiates multiple GNN training processes in parallel on both the CPU and the GPU.
By allocating training processes on the CPU to perform GNN training collaboratively with the GPU, the proposed protocol improves the platform resource utilization and reduces the CPU-GPU data transfer overhead.
Since the performance of a CPU and a GPU varies, we develop a novel load balancer that balances the workload dynamically between CPUs and GPUs during runtime.
We evaluate our protocol using two representative GNN sampling algorithms, with two widely-used GNN models, on three datasets. 
Compared with the Standard training protocol adopted in the state-of-the-art GNN frameworks, our protocol effectively improves resource utilization and improves the overall training time.
On a platform where the GPU moderately outperforms the CPU, our protocol speeds up GNN training by up to 1.41$\times$.
On a platform where the GPU significantly outperforms the CPU, our protocol speeds up GNN training by up to 1.26$\times$.
{Our protocol is open-sourced and can be seamlessly integrated into state-of-the-art GNN frameworks and accelerate GNN training.
Our protocol particularly benefits those with limited GPU access due to its high demand.}
\end{abstract} 

\maketitle

\section{Introduction}\label{sec:intro}
\begin{figure*}[t]
    \centering
    \includegraphics[width=13.5cm]{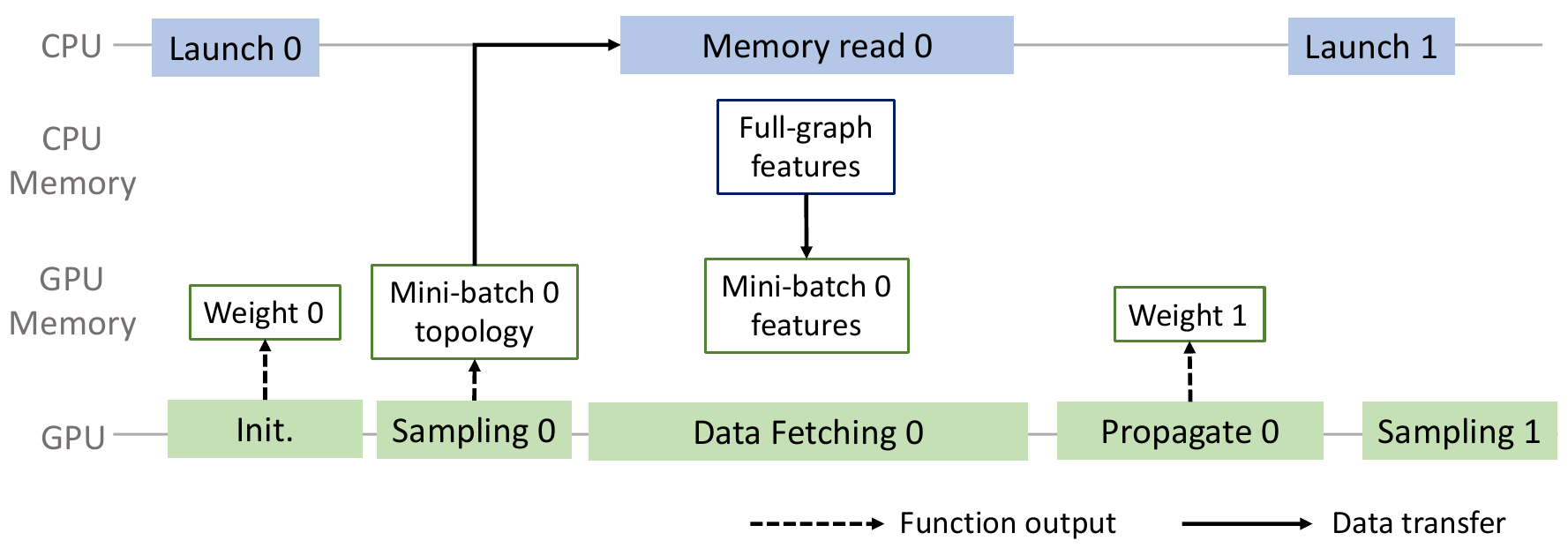}
    \caption{Standard training protocol used in state-of-the-art GNN frameworks}
     \label{fig:old_protocol}
\end{figure*}

Graph Neural Network (GNN) is an emerging type of Machine Learning model that can extract useful information from graph-structured data.
GNNs are widely used in various applications such as Electronic Design Automation~\cite{gnn-eda,eda2}, molecular property prediction \cite{yang_li_2023, graphsage}, and social recommendation system \cite{recommend1, recommend2,DLRM_GPU}, where the input is often a large-scale graph with over billion edges.
Training a GNN model on these real-world graphs involves a high volume of data communications and computations.
While state-of-the-art CPUs and GPUs feature high compute power and memory bandwidth, existing GNN training protocols adopted in popular frameworks like PyTorch-Geometric (PyG) \cite{pyg} and Deep Graph Library (DGL) \cite{dgl} cannot efficiently utilize the available platform resources;
therefore, large-scale GNN training is time-consuming, taking hours or even days \cite{dorylus}.
We conduct a detailed analysis of state-of-the-art GNN frameworks and observe several inefficiencies in the existing Standard GNN training protocols.
First, Standard GNN training protocols (shown in Figure~\ref{fig:old_protocol}) offload most of the workload to the GPUs, leaving the CPUs mostly idle.
Such a task allocation is sub-optimal as state-of-the-art CPUs offer comparable GNN training performance with GPUs.
For example, the epoch time of training a three-layer GCN~\cite{gcn} model on the ogbn-products~\cite{ogb} dataset takes 10 seconds using an Intel Xeon 8280 with only 28 cores \cite{distgnn}, and takes 5 seconds using a high-end NVIDIA V100 GPU \cite{gnnlab}.
This suggests that CPUs should be incorporated in GNN training due to their potential to enhance overall performance.
In addition, existing GNN training protocols adopt a coarse-grained task scheduling, alternating between memory-intensive and compute-intensive operations. 
This approach leads to sub-optimal resource utilization: the memory bandwidth is under-utilized during compute-intensive tasks, and the compute cores are under-utilized during memory-intensive tasks.

Motivated by the above challenges, we propose a novel Unified CPU-GPU training protocol that improves the resource utilization of GNN training on a CPU-GPU platform.
The Unified CPU-GPU protocol defines the CPU-GPU interaction to perform GNN training collaboratively.
Note that our protocol not only supports platforms with a single CPU and single GPU, but also platforms with multiple CPUs and multiple GPUs.
{Our protocol instantiates GNN training processes on both the CPUs and the GPUs, as opposed to offloading all the workload to the GPUs.
Allocating GNN training processes on the CPUs to share a portion of the workload leads to several advantages:
(1) It allows our protocol to exploit the compute cores of CPUs to execute GNN operations, leading to higher platform resource utilization than the Standard training protocol, and reduces the workload on the GPUs;
(2) It reduces the data transferred to the GPU via the PCIe channel, which often causes significant overhead;
(3) It reduces the memory usage on the GPU global memory, and the memory space that becomes available can be used for data caching, which further reduces the expensive PCIe data transfer. 
Furthermore, instantiating multiple GNN training processes improves memory bandwidth utilization because it overlaps the communications with computations across the processes.
Since the performance of the CPU and GPU for training a GNN model varies, we develop a Dynamic Load Balancer to ensure the workload is balanced; otherwise, one platform can easily become the system bottleneck and lead to performance degradation.}
{As GPUs often face high demand from multiple users, and the CPUs are becoming increasingly powerful, our protocol exploits the CPU resources, which are under-utilized by the Standard training protocol, to improve GNN training performance.}
% The Dynamic Load Balancer is a lightweight solution that handles load balancing during runtime. 
% To balance the workload between the CPUs and the GPUs, we propose a lightweight resource manager that automatically collects runtime information to fine-tune the workload distribution and perform on-the-fly resource allocation.
% Furthermore,
% A device (a CPU or a GPU) can start to perform propagation as soon as its process finishes fetching the required data, while other processes keep fetching data from the CPU memory.
% This overlaps , and improves the memory bandwidth utilization, resulting in improved performance.
% Finally, we conduct detailed profiling on the CPU design and identify that a noticeable amount of time is spent on inefficient memory allocation;
% we propose a simple yet effective solution that improves the memory allocation efficiency, subsequently improving the performance of the hybrid CPU-GPU training.
Note that none of the proposed optimizations alter the GNN training semantics, meaning that training with the Unified CPU-GPU protocol leads to the same model accuracy and convergence rate as the Standard training protocol.
Our key contributions are:
\begin{itemize}
    \item We conduct a detailed analysis of the state-of-the-art GNN frameworks to identify the inefficiencies in the Standard GNN training protocol. 
    \item We propose a novel Unified CPU-GPU protocol for GNN training that can effectively improve the utilization of both the compute resources and memory bandwidth.
    \item {We develop a novel Dynamic Load Balancer that effectively balances the workload between the CPUs and the GPUs dynamically during runtime.} 
    % \item Our library instantiates multiple GNN training processes to increase computation-communication overlapping, which improves memory bandwidth utilization and overall training performance.
    \item {We evaluate our work using various CPU-GPU platforms: on a platform where the GPU moderately outperforms the CPU, our protocol accelerates GNN training by up to 1.41$\times$; on a platform where the GPU significantly outperforms the CPU, our protocol accelerates GNN training by up to 1.26$\times$.} 
    \item Our protocol is open-sourced\footnote{\url{https://github.com/jasonlin316/A-Unified-CPU-GPU-Protocol-for-GNN-Training}} and can seamlessly integrate into existing GNN frameworks such as PyG and DGL to improve GNN training performance.
\end{itemize}

\section{Background}
\subsection{Graph Neural Networks}
% Brief introduction of GNN. Include two examples: GCN and GraphSAGE.

% 1. Take Structure Information and Node/Edge Features as input
% 2. Output Graph representations: node-wise, edge-wise, graph-wise. Edge and graph-wise representations are often based on node-wise representation. For simplicity, we only consider node-wise representation in this work.
% 3. GNN includes Propagation and Transformation, which are related to SpMM and GEMM, respectively.

Given a graph $G=(V, E)$ with $N = |V|$ nodes and $|E|$ edges.
Let $A \in \mathbb{R}^{N \times N}$ be the adjacency matrix with an entry $(i, j)$ equal to 1 if there exists an edge between node $i$ and $j$, otherwise 0.
For each node $v \in V$, $N(v)$ is the set of neighbors of $v$, and $x_v$ is a $F$-dimensional feature vector associated with $v$. 
We use $H^0 \in \mathbb{R}^{N \times F}$ to denote the input feature matrix of $V$.
A Graph Neural Network (GNN) is a neural network that operates on graph-structured data with node-related features as input. 
By aggregating feature information through graph structure and transforming the features into $d$-dimensional latent space, GNNs are able to output representations of nodes containing higher-order neighbor information. 
{GNN models~\cite{gcn, gin, gat, graphsage} follow the above two operations (i.e., aggregation and transformation) and can be described using the Message-Passing paradgim~\cite{message-passing}.} 
% Two fundamental operators in a Message-Passing layer are Sparse Matrix-Matrix Multiplication (SpMM) and General Matrix-Matrix Multiplication (GEMM), which correspond to the aggregation and transformation operations, respectively.
We list two representative GNN models as examples:
\begin{itemize}
    \item
        GCN~\cite{gcn} is one of the most widely used GNN models. The $l$-th layer of GCN can be defined as follows:
        \begin{equation}
        H^{(l)}=\sigma\left(\hat{A} H^{(l-1)} W^{(l)}\right),
        \end{equation}
        $W^{(l)}$ and $H^{(l)}$ indicates the weight matrix and feature matrix of layer $l$, respectively.
        $\hat{A}$ is the normalized and regularized adjacency matrix and $\sigma(\cdot)$ is the activation function. 
        Multiplying $\hat{A}$ with $ H^{(l-1)}$ aggregates the feature information, while multiplying $\hat{A} H^{(l-1)}$ with $W^{(l)}$ transforms the aggregated features into a $d$-dimensional latent space.
    \item
        GraphSAGE~\cite{graphsage} added a self-concatenation upon the GCN layer. The $l$-th layer of GraphSAGE can be defined as follows:
        \begin{equation}
        H^{(l)}=\sigma\left(H^{(l-1)} W_1^{(l)} + \hat{A} H^{(l-1)} W_2^{(l)}\right).
        \end{equation}
\end{itemize}
{Our Unified GNN training protocol supports all GNN models that follow the Message-Passing paradigm, including other widely used models such as GIN~\cite{gin} and GAT~\cite{gat}.}

\begin{figure}[t]
    \centering
    \includegraphics[width=7cm]{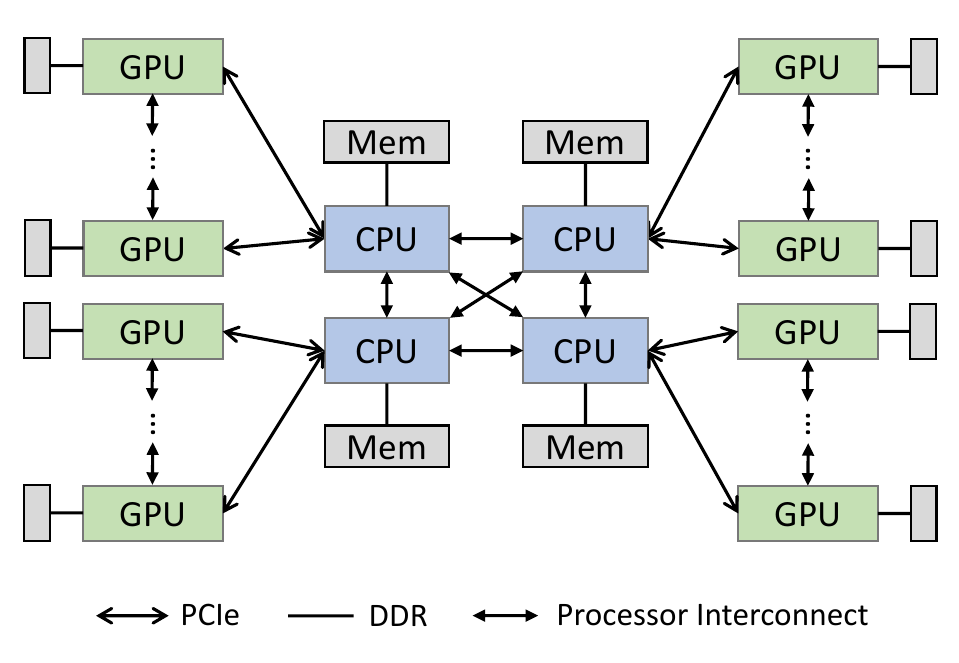}
    \vspace{-0.2cm}
    \caption{Target CPU-GPU platform}
     \label{fig:platform}
\end{figure} 

\subsection{Mini-batch SGD GNN Training}\label{sec:algo}
% Explain what the mini-batch GNN training is. Two examples: Neighbor Sampler and ShaDow Sampler.

% Introduction of full-batch training and mini-batch SGD training.
GNNs are initially trained using full gradient descent \cite{gcn}, which takes the whole adjacency matrix $A$ and all nodes' features $H^{0}$ as input. However, full gradient descent suffers from unacceptable memory costs on large graphs and requires more epochs to converge since the model is updated only once per epoch.
To address these issues, mini-batch SGD has been proposed for GNN training \cite{graphsage}. Unlike full gradient descent, mini-batch SGD does not require computing the gradient for all the nodes in the graph at once. Instead, it samples a subgraph based on a batch of node indices and applies the GNN model on top of the subgraph to calculate the gradient.
Shown in Figure \ref{fig:old_protocol}, to perform mini-batch training, the full graph can be stored in the CPU memory.
After a mini-batch is sampled, the CPU transfers the node embeddings of the sampled nodes to the GPU memory for model propagations (i.e., data fetching).
Let $\mathcal{B} \subseteq V$ with size $b=|\mathcal{B}|$ denote a batch of vertices. The update in every SGD step is based on the following gradient estimation:
\begin{equation}
\frac{1}{|\mathcal{B}|} \sum_{i \in \mathcal{B}} \nabla \mathcal{L}\left(y_{i}, h_i^{(L)}\right),
\end{equation}
where $h_i^{(L)}$ is the $i$-th row of $H^{(L)}$ with ground truth label $y_{i}$.
{The mini-batch can be obtained via various approaches \cite{clustergcn,graphsaint,graphsage,shaDow}; we list two representative sampling algorithms as examples:}
\begin{itemize}
    \item
        Neighbor Sampling \cite{graphsage} randomly selects neighbors for each node and enforces a predefined budget on the sample size for each layer. For a GNN model with $L$ layers, it samples a subgraph within $L$-hops for each root node.
    \item
        ShaDow K-Hop Sampling \cite{shaDow} is a variant of Neighbor Sampling. It applies an $L$-layer GNN (with arbitrary depth $L$) on top of a localized $L'$-hop subgraph associated with each root node, where $L$ is often set to 5 and $L'$ is set to 2. Each shallow subgraph is sampled using Neighbor Sampling.
\end{itemize}
% Introduce synchronous SGD, and clarify there is no sample dependencies between different batch node.
We adopt synchronous SGD \cite{sgd, ddp} to train GNNs on multiple devices, as it is widely used for multi-GPU training \cite{p3,pagraph} and distributed training across multiple machines \cite{distdgl,distgnn}; most importantly, it is algorithmically equivalent to training with a larger mini-batch on a single device, as long as the sample gradient calculation $\nabla \mathcal{L}\left(y_{i}, h_i^{(L)}\right)$ is independent between different samples $i$. 
For synchronous SGD training, multiple sub-mini-batches are first sampled and assigned to different devices. 
Next, forward propagations are performed on each device based on the original GNN algorithm. Finally, gradients are gathered and averaged across devices to calculate the actual mini-batch gradient for model updates.
% Training with synchronous SGD on multiple devices is algorithmically equivalent to training with a larger mini-batch on a single device, as long as the sample gradient calculation $\nabla \mathcal{L}\left(y_{i}, h_i^{(L)}\right)$ is independent between different samples $i$. 
% However, subgraph sampling methods \cite{clustergcn, graphsaint} and layer-dependent sampling methods \cite{asgcn, ladies} violate the independence requirements between batch nodes. Therefore, there is no guarantee of accuracy for these methods when the sub-batch size assigned to each device varies.
% To preserve the semantics of original GNN training algorithms, we use two node-wise sampling methods in this work that comply with the independence requirements:

\begin{figure}[t]
    \centering
    \includegraphics[width=7.5cm]{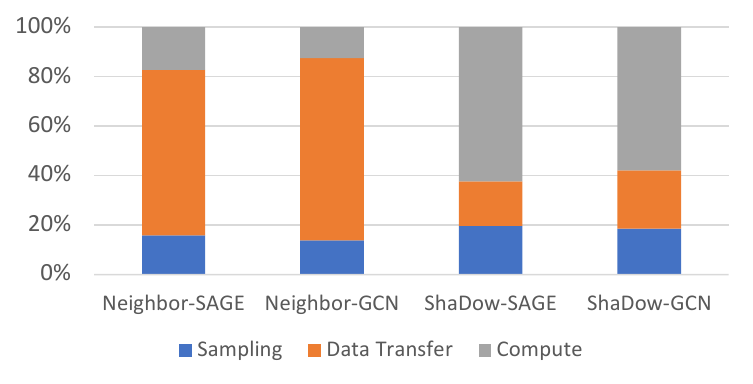}
    \vspace{-0.2cm}
    \caption{Training time breakdown of existing GNN library for various sampling algorithms and GNN models}
     \label{fig:dgl_breakdown}
\end{figure} 

\subsection{Target CPU-GPU Platform}\label{sec:cpu_gpu}
Figure \ref{fig:platform} shows the target CPU-GPU platform. 
The platform consists of multiple CPUs and multiple GPUs. 
The CPUs and GPUs are connected through PCIe.
The CPUs are connected to other CPUs via processor interconnects such as UPI;
similarly, the GPUs are connected to other GPUs via GPU-specific processor interconnects such as Nvidia NVLink \cite{nvlink}. 
Each CPU and GPU is connected to a memory via DDR memory channel. 
State-of-the-art libraries like CUDA and SYCL provide users with a unified view of the CPU and GPU memory system, accessing both the CPU main memory and GPU global memory as a shared memory space. 
However, it’s important to note that the shared memory space actually consists of various memory channels, such as PCIe, DDR, and UPI, each with a different bandwidth. Thus, to achieve high performance, it is essential to exploit optimizations to mitigate the overhead of accessing channels with low memory bandwidth (Section \ref{sec:malloc}).

% While there are multiple DDR memories, some are connected to CPU and some to GPU, our protocol (Section \ref{sec:protocol}) regards them as a unified shared memory space that can be accessed by all the CPUs and GPUs on the platform;
% such function (i.e., creating shared memory space across CPUs and GPUs) is supported by state-of-the-art programming libraries, for example, Unified Shared Memory (USM) \cite{usm} in SYCL and Unified Virtual Address (UVA) \cite{uva} in CUDA.

\begin{figure*}[t]
    \centering
    \includegraphics[width=15.5cm]{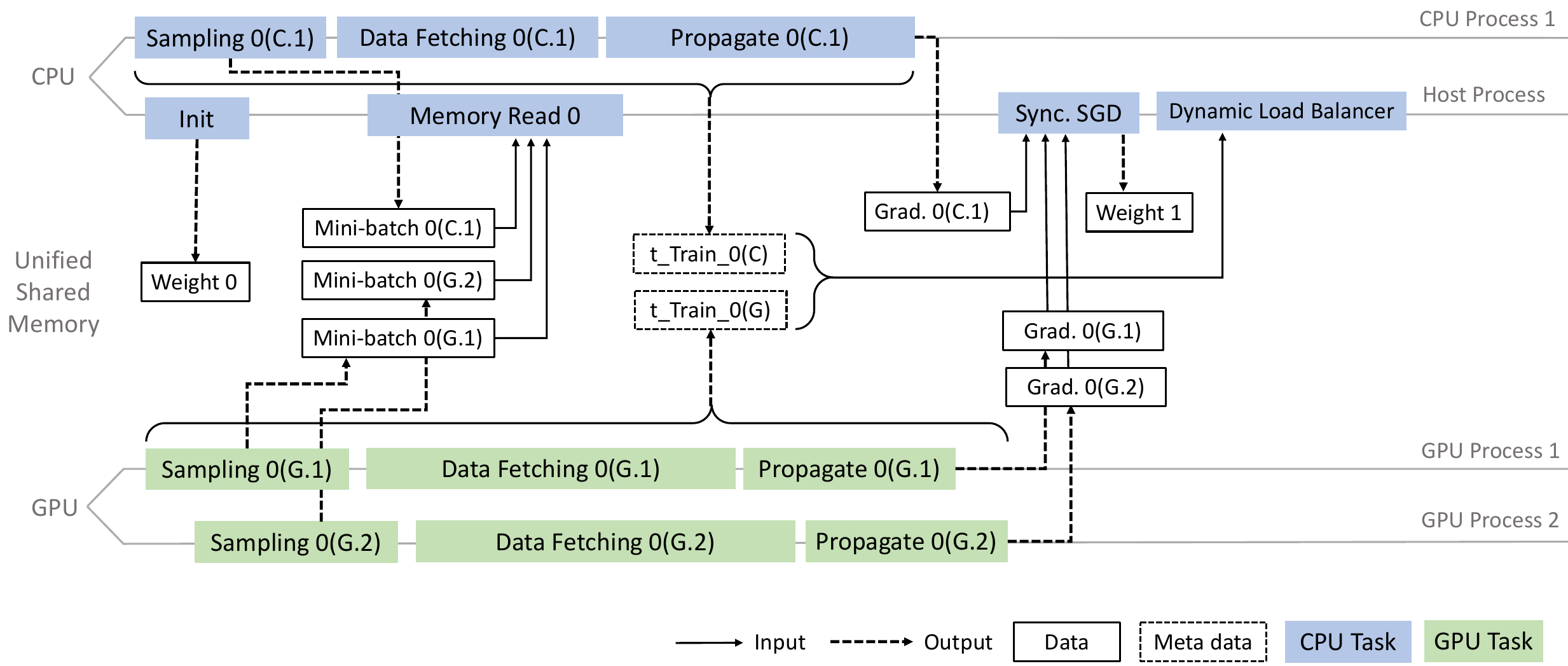}
    \vspace{-0.1cm}
    \caption{Unified CPU-GPU training protocol}
     \label{fig:new_protocol}
\end{figure*} 

\begin{figure}[t]
    \centering
    \includegraphics[width=7cm]{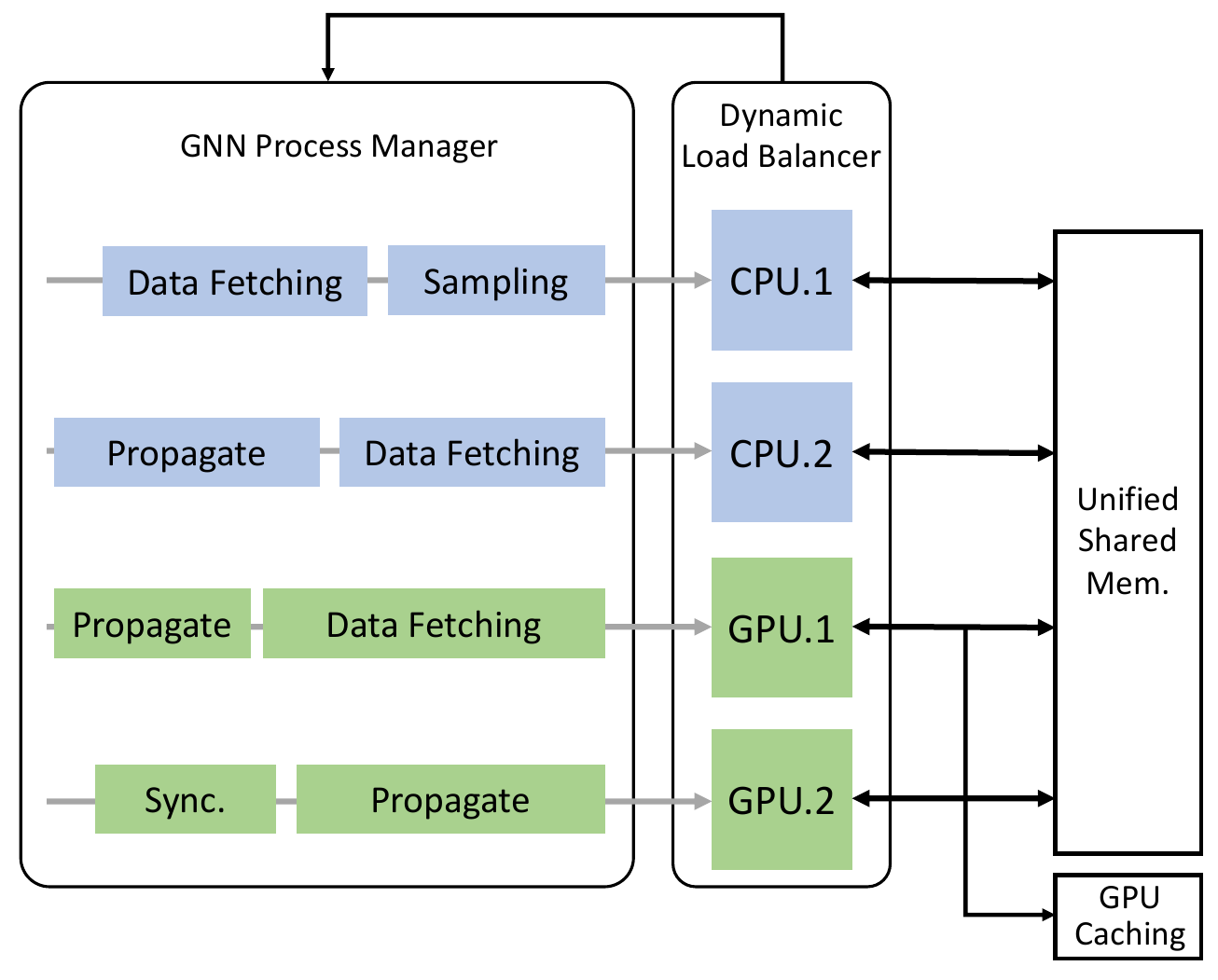}
    \caption{System Overview}
     \vspace{-0.2cm}
     \label{fig:overview}
\end{figure} 

\section{Unified CPU-GPU Protocol for GNN Training}\label{sec:protocol}
To better understand the inefficiencies in the Standard GNN training protocols adopted by the state-of-the-art GNN frameworks, we use Deep Graph Library (DGL) \cite{dgl} as an example and perform profiling on the MAG240M~\cite{hu2021ogblsc} dataset with several sampling algorithms and GNN models.
As shown in Figure \ref{fig:dgl_breakdown}, the runtime of GNN training with the Neighbor Sampling algorithm is dominated by the data transfer overhead between CPU and GPU.
On the other hand, under the ShaDow K-Hop Sampling algorithm, the system bottleneck shifts to computation.
Both cases suggest that using CPU in GNN training can improve performance
by handling a portion of the workload.
By doing so, both the CPU-GPU data traffic and the workload on the GPU can be reduced.
{Furthermore, our profiling results show that the CPU is heavily under-utilized: the average CPU utilization for GNN training is below 3\%. This suggests a potential for assigning some of the workload to the CPU platform.}

We propose a novel Unified CPU-GPU Protocol, which defines the CPU-GPU interaction to perform GNN training, collaboratively.
Our protocol supports platforms with multiple CPUs and GPUs. 
For simplicity, in Figure \ref{fig:new_protocol}, we illustrate our protocol using a platform with one CPU and two GPUs as an example.
We first explain the meaning of the suffix in each function block and data block: the first number indicates the number of the training iteration;
the second alphabet in the parentheses indicates the device type (i.e., CPU or GPU) that holds the data or executes the function;
the last number indicates the device number (e.g., GPU 1, GPU 2).
For example, \texttt{Sampling 0(G.1)} indicates the sampling operation performed in training iteration 0 on GPU 1;
Similarly, \texttt{Grad. 0(C.1)} indicates the local gradient generated in training iteration 0 on CPU 1.

Our protocol adopts the shared memory model for data exchange between the CPUs and the GPUs, i.e., it creates a unified shared memory space as mentioned in Section \ref{sec:cpu_gpu}.
The CPU platform consists of two types of processes: the Host Process and the CPU Process.
The Host Process is responsible for launching GNN tasks, setting barriers for synchronization, and load balancing.
The CPU Process is responsible for executing GNN operations, such as sampling and model propagation (i.e., forward and backward propagations), on the CPU platform.
Similar to the Standard training protocol depicted in Figure \ref{fig:old_protocol}, the sampling operation constructs subgraphs for each mini-batch. The node and edge sets of these subgraphs are then used for data fetching to load the corresponding feature vectors from the full graph for model propagation.
We omit the data blocks of the full-graph features and mini-batch features in Figure \ref{fig:new_protocol} for simplicity.
The CPU Processes allow our protocol to exploit CPU compute cores to perform GNN training, which reduces the workload on the GPU.
Then, the GPU Processes execute the GNN operations on the GPU platform.
After the GNN operations are performed, each CPU and GPU Process generates a local gradient; the local gradients are then gathered to perform a synchronous stochastic gradient descent (Sync. SGD) \cite{sgd} to update the GNN model globally.
In addition to model updates, the Host Process collects the runtime information for the Dynamic Load Balancer.
The Dynamic Load Balancer performs workload-aware sub-batch assignments among the CPUs and the GPUs based on a certain ratio;
if the ratio leads to workload imbalance, the Dynamic Load Balancer adjusts the ratio for the next epoch using the collected runtime information.
We explain the mechanism of Dynamic Load Balancer in detail in Section \ref{sec:manager}.

\section{System Design}\label{sec:opt}

We depict the overview of our Unified CPU-GPU training system in Figure \ref{fig:overview}.
Our system consists of several building blocks to execute the Unified CPU-GPU protocol.
It is worth noting that none of the optimization applied in our system alters the GNN training semantics, meaning that our system does not trade off model accuracy or convergence rate for speed.
% We describe each building block of our system in detail in the following subsections.

%emphasize interleaving, fine-grained scheduling
\subsection{GNN Process Manager}\label{sec:multi_process}
To perform Unified CPU-GPU training, we develop a GNN Process Manager that allocates GNN training processes on both the CPU and the GPU.
On the CPU platform, two processes are running, one being the host process and the other being the CPU process, and on the GPU platform, $n$ process is running simultaneously, where $n$ is the number of GPUs on the platform.
Compared with only allocating GNN training processes on the GPU, this leads to performance improvement because allocating GNN processes on the CPU platform reduces the workload on the GPU, and also reduces the amount of data that needs to be transferred to the GPU via the PCIe.
Furthermore, instantiating multiple GNN training processes increases memory bandwidth utilization by overlapping the computations with communications.
We show an example in Figure \ref{fig:overview}:
while some processes perform data fetching (communications), others perform mini-batch sampling or model propagation (computations).
Note that such scheduling does not alter the GNN training semantic because synchronous SGD only requires each process to synchronize at the end of each iteration, so it is valid to have processes executing different tasks (e.g., sampling, propagation) during training.

% Existing GNN library \cite{dgl} adopts coarse-grained scheduling, which first fetches all the required data, and then distributes the data to each device for computation.
% This leads to low bandwidth utilization since the memory bandwidth is under-utilized during the computations.
% To improve bandwidth utilization, our system features a GNN Process Manager that instantiates multiple GNN training processes to perform GNN operations;
% this allows the system to overlap the communications with computations, increasing bandwidth utilization.

\subsection{Dynamic Load Balancer} \label{sec:manager}
{To train GNN with multiple devices, such as multiple GPUs, the state-of-the-art runtime system \cite{ddp} assigns the same number of mini-batches to each device by default.
% Such an assignment is based on the assumption that the performance of each device is identical.
However, since the performance of CPU and GPU can be different, assigning the same number of mini-batches to the CPU and the GPU leads to workload imbalance;
therefore, we develop a load balancer to distribute the mini-batches.}
{A straightforward way to achieve load balancing on the CPU-GPU platform is to distribute the mini-batches to the CPU and GPU based on their relative performance;
the relative performance can be derived by running one epoch of training on both platforms and measuring the execution time respectively.
For example, assume the GPUs are $2\times$ faster than the CPUs; to balance the workload, the system can assign $1/3$ of the mini-batches to the CPUs and $2/3$ of the mini-batches to the GPUs.
We refer to this balancing mechanism as "Static Load Balancing," as the number of mini-batches is statically assigned to the CPUs and GPUs.
We define the \textit{workload ratio} as the workload assignment between the CPU and the GPU; the workload ratio of the above example is 1:2.  
If the current workload ratio leads to workload imbalance, the Load Balancer adjusts the workload ratio accordingly in the next iteration.
Specifically, the Load Balancer collects the execution time of CPU and GPU, and recalculates the workload ratio to balance the workload.
% For example, when the CPU is bottlenecking the system, the Load Balancer decreases the workload on the CPU by half of its original workload;
% if the CPU becomes faster than the GPU after the adjustment, the Load Balancer increases the workload on the CPU by a quarter of its original workload, and so on and so forth.
Static Load Balancing relies on the assumption that the workload of training on each mini-batch is similar (i.e., uniform workload distribution), so the number of mini-batches can be used to estimate the total workload;
however, this is not true for some datasets, where the mini-batches have a highly skewed workload distribution.
In such cases, workload balancing cannot be achieved merely by considering the number of mini-batches.}

{We develop a Dynamic Load Balancer to address the issue that the mini-batches can have a highly skewed workload distribution.
Given a sampling algorithm and a dataset, the Dynamic Load Balancer estimates the workload of each mini-batch in advance during the pre-processing stage;
the estimation is performed by running the sampling algorithm before the actual GNN model training and calculating the total number of aggregations that need to be performed using the computational graph of the mini-batches.
The estimation process is a one-time cost that can amortized.
Afterward, similar to Static Load Balancing, the Dynamic Load Balancer estimates the workload ratio that balances the workload between the CPU and the GPU.
Assume the GPUs are $2\times$ faster than the CPUs.
Instead of statically assigning $1/3$ of the mini-batches to the CPU, the Dynamic Load Balancer sorts the mini-batches by their estimated workload and assigns a certain number of mini-batches that account for $1/3$ of the total workload to the CPU.
This workload-aware sub-batch assignment leads to better load balancing than statically assigning a fixed number of mini-batches to the CPUs and GPUs.
Since the number of mini-batches assigned to the CPUs and GPUs is adjusted dynamically during runtime, we refer to this mechanism as "Dynamic Load Balancing."
After training for one epoch, the Dynamic Load Balancer collects the runtime information to evaluate the workload ratio.
If the workload is imbalanced, the Dynamic Load Balancer recalculates the workload ratio using the collected runtime information, and adjusts the ratio accordingly.}

\subsection{GPU Feature Caching}\label{sec:malloc}
% While many profiling works have been done to understand the performance bottlenecks in GNN training on GPU platforms \cite{gpu_profile1, gpu_profile2, gpu_profile3}, little work has been done on the CPU platform;
% however, it is equally important to understand the CPU implementation for hybrid CPU-GPU training.
% We conduct a detailed analysis (Section \ref{sec:cpu_prof}) on the CPU implementation and notice that a major portion of the GNN training time is spent on memory allocation. 
% This is because the serial memory allocator used by default does not scale well for multi-threaded programs \cite{hoard_malloc}, causing a large overhead.
% To overcome the inefficiency of memory allocation, our system replaces the default memory allocator with a parallelized multi-threaded memory allocator, for example, Hoard \cite{hoard_malloc} and TCMalloc \cite{tcmalloc}; this simple yet effective change results in up to $1.8\times$ training performance compared with using the serial memory allocator.

% The Unified CPU-GPU protocol assigns a portion of the mini-batches to the CPU, rather than offloading all the mini-batches to the GPU. This reduce the amount of node and edge features transferred to the GPU, as well as reduces the usage of the GPU global memory.

% Dynamic load balancing not only achieves a smaller synchronization time than static load balancing between the CPU and GPU processes but also increases the workload ratio of the CPU.
By assigning a portion of the mini-batches to the CPU, the Unified CPU-GPU protocol reduces the amount of CPU-GPU data communication, as well as the usage of the GPU global memory;
the additional memory capacity allows us to perform GPU feature caching, which stores a set of frequently accessed feature vectors in the GPU global memory.
Such a mechanism reduces the expensive CPU-GPU communication overhead via PCIe during the data fetching operation (see Figure \ref{fig:new_protocol} for detail).
Specifically, when loading the feature vectors of a mini-batch from the unified shared memory, if a vector resides in the GPU global memory, it eliminates the need for memory access over the PCIe.
In this work, we integrate a GPU feature caching approach, as proposed by HugeCTR~\cite{hugeCTR}, into our system. 
This approach utilizes the least recently used cache replacement strategy (LRU) to retain the most recently accessed features on the GPU.
Note that for large-scale datasets, most feature vectors are stored in the CPU main memory, which has a greater capacity (several terabytes) than the GPU global memory (several gigabytes).
This storage disparity results in high communication overhead, as most data must be fetched from the CPU main memory. 
Such a characteristics makes GPU feature caching particularly useful, as it reduces the data traffic of fetching feature vectors from the CPU main memory.

% Note that GPU caching is a mechanism that manually stores some data in the GPU global memory, which is different from traditional caching, where the operating system automatically stores some data in the cache.
% By caching nodes and edges in the GPU global memory, our system reduces the expensive CPU-GPU communication via PCIe.

% Note that data stored in the unified shared memory can be either on the CPU DDR memory or GPU global memory, and there is no performance boost by using GPU caching if the data is already on the GPU global memory;
% however, the majority of data are stored in CPU memory, which has a larger capacity (several terabytes) than the GPU global memory (several gigabytes), thus, GPU caching remains beneficial.

\section{Experiments}\label{sec:exp}
\newcolumntype{Y}{>{\centering\arraybackslash}X}

\subsection{Experiemental Setup}
\subsubsection{Environment}\label{sec:env}
We conduct our experiments on two CPU-GPU platforms. 
{Both platforms are equipped with the same CPU, which is a two-socket Intel Xeon Gold 6326; the CPU and GPU are connected via PCIe.
Platform 1 consists of a high-end NVIDIA A100 GPU with the Multi-Instance GPU (MIG) \cite{MIG} technology enabled.
MIG is a technology that allows GPU resources to be shared efficiently on the cloud by partitioning the GPU into multiple instances.
Such a setup is becoming increasingly important in real life as data center GPUs like A100 are often shared by multiple users.
In addition, the MIG setup aligns with our protocol, which aims to improve performance under scenarios with limited GPU access.
We adopt the 3g.20gb configuration, which partitions the A100 into two instances.
Note that even with MIG enabled, A100 (3g.20gb) still offers superior peak performance compared with the CPU platform (Table \ref{tab:spec}).
Platform 2 consists of a workstation NVIDIA RTX A5000 GPU.}
We list the detailed specifications of the CPUs and GPUs used in our experiments in Table \ref{tab:spec}.
We implement our design using Python v3.8, PyTorch v2.0.1 with CUDA 11.7, and Deep Graph Library (DGL) v1.1.2.
{To profile the CPU utilization and memory bandwidth, we use the Intel VTune Profiler \cite{vtune}; to profile the GPU utilization, we use the NVIDIA System Management Interface (nvidia-smi) \cite{nvsmi}.}

\subsubsection{Sampler, GNN Models, and Datasets}
To evaluate our system, we use three large-scale datasets: Reddit \cite{graphsaint}, ogbn-products \cite{ogb}, and MAG240M \cite{hu2021ogblsc}.
% Note that MAG240M is a heterogeneous graph with various types of nodes and edges;
% we convert MAG240M to a homogeneous graph for our experiments.
We choose two widely-used GNN models: GCN \cite{gcn} and GraphSAGE \cite{graphsage}.
To generate the mini-batches, we choose two representative GNN sampling algorithms: the Neighbor Sampler \cite{graphsage} and the ShaDow Sampler \cite{shaDow}.
The Neighbor Sampler constructs a three-layer model with a sampling size [15, 10, 5]; the ShaDow sampler first produces a localized three-hop subgraph with a sampling size [15, 10, 5] and then constructs a five-layer model on the subgraph (i.e., $L'=3$, $L=5$; details in Section \ref{sec:algo}). 
The hidden feature size is set to 128.
We set the mini-batch size as 4096 for the Reddit and ogbn-products datasets, and set the mini-batch size as 1024 for MAG240M;
this is because the MAG240M input feature size is larger than the other two datasets, and setting a batch size of 4096 leads to an out-of-memory error.
We list the details of the three datasets in Table \ref{tab:dataset}, which includes the number of nodes, edges, and the feature size of the input $f_0$  and output layer $f_L$.

\begin{table}[t]
\caption{Specifications of the platforms }
\begin{adjustbox}{max width=1\columnwidth}
\renewcommand{\arraystretch}{0.9}
\begin{tabular}{c|c|cc}
\toprule
Platforms           & {CPU}              & {Platform 1}    & {Platform 2}              \\ \midrule
\multirow{2}{*}{Devices} & Intel                & NVIDIA                         & NVIDIA               \\
                         & Xeon 6326            & A100 (3g.20gb)                      & RTX A5000            \\ \midrule \midrule
% \# of Units              & 2                    & 1                    & 2                    & 4                    \\
Frequency                & 2.9 GHz              & 1405 MHz                   & 2000 MHz             \\
Peak Performance         & 537 GFLOPS           & 8.36 TFLOPS                & 27.8 TFLOPS          \\
Last-level Cache         & 24 MB L3             & 20 MB L2                     & 6 MB L2              \\
Memory Bandwidth         & 171 GB/s             & 778 GB/s                      & 768 GB/s             \\
\bottomrule
\end{tabular}
\end{adjustbox}
\label{tab:spec}
\end{table}

\begin{small}
\begin{table}[t]
\caption{Specifications of the datasets}
\renewcommand{\arraystretch}{0.9}
\begin{tabularx}{\columnwidth}{YYYcc}
\toprule
Datasets        &  Vertices &  Edges   & $f_0$ & $f_L$ \\ \midrule \midrule
Reddit & 232,965    & 11,606,919 & 602            & 41           \\ 
ogbn-products   & 2,449,029        & 61,859,140    & 100            & 47             \\ 
MAG240M         & 244,160,499    & 1,729,762,391 & 768            & 153            \\ \bottomrule
\end{tabularx}
\label{tab:dataset}
\end{table}
\end{small}

\begin{lstlisting}[caption={Single GPU training with DGL library},label=lst:code2,float=t]
def train(...):
  device = torch.device("cuda")
  model = GNN(...).to(device)
  loader = dgl.dataloading.DataLoader(
    g,
    train_idx.to(device),
    dgl.dataloading.NeighborSampler(...),
    device=device,
    use_uva=device.type == "cuda")
  opt = torch.optim.Adam(...)
  for epoch in range(runs):
    _train(loader, model, opt)
    ...
\end{lstlisting}

\subsection{System Implementation}\label{sec:program_mod}
Our protocol can be seamlessly incorporated into existing GNN frameworks.
In Listing~\ref{lst:code2}, we show an example program that performs GNN training on a single GPU platform using DGL.
In Listing~\ref{lst:code3}, we highlight the modifications to train with our Unified CPU-GPU training protocol.
We only show an example using DGL because the modifications are similar to incorporating our protocol into PyG.
The program modifications include (1) applying PyTorch 
\begin{lstlisting}
def train(...):
\end{lstlisting}
% \vspace{-\baselineskip}
\reducespace
\begin{lstlisting}[backgroundcolor=\color{codebg2}]
  torch.distributed.init_process_group("gloo", ...)
\end{lstlisting}
% \vspace{-\baselineskip}
\reducespace
\begin{lstlisting}[backgroundcolor=\color{codebg1}]
  device = get_device()
\end{lstlisting}
% \vspace{-\baselineskip}
\reducespace
\begin{lstlisting}
  model = GNN(...).to(device)
\end{lstlisting}
% \vspace{-\baselineskip}
\reducespace
\begin{lstlisting}[backgroundcolor=\color{codebg2}]
  model = torch.DistributedDataParallel(model)
\end{lstlisting}
% \vspace{-\baselineskip}
\reducespace
\begin{lstlisting}
  loader = dgl.dataloading.DataLoader(
    g,
\end{lstlisting}
% \vspace{-\baselineskip}
\reducespace
\begin{lstlisting}[backgroundcolor=\color{codebg1}]
    UnevenDDPIndices(train_idx.to(device)),
\end{lstlisting}
% \vspace{-\baselineskip}
\reducespace
\begin{lstlisting}
    dgl.dataloading.NeighborSampler(...),
    device=device,
    use_uva=device.type == "cuda")
  opt = torch.optim.Adam(...)
\end{lstlisting}
\reducespace
\begin{lstlisting}[backgroundcolor=\color{codebg2}]
  balancer = DynamicLoadBalancer()
\end{lstlisting}
% \vspace{-\baselineskip}
\reducespace
\begin{lstlisting}
  for epoch in range(runs):
\end{lstlisting}
% \vspace{-\baselineskip}
\reducespace
\begin{lstlisting}[backgroundcolor=\color{codebg1}]
    profile = unified_train(balancer.config(), 
        _train, args=(loader, model, opt))
\end{lstlisting}
% \vspace{-\baselineskip}
\reducespace
\begin{lstlisting}[backgroundcolor=\color{codebg2}]
    balancer.update(profile)
\end{lstlisting}
% \vspace{-\baselineskip}
\reducespace
\begin{lstlisting}[caption={Unified CPU-GPU Training with our protocol \\ \textcolor{codebg2}{$\blacksquare$}: newly added lines. \textcolor{codebg1}{$\blacksquare$}: modified lines},label=lst:code3]
    ...
\end{lstlisting}
\texttt{DistirbutedDataParallel} wrapper to enable multi-process GNN training, (2) applying the \texttt{unified\_train} wrapper to enable Unified CPU-GPU training, and (3) launching the Dynamic Load Balancer.
The Dynamic Load Balancer takes the runtime information as input, and adjusts the workload distribution between CPU and GPU on the fly.
However, the native \texttt{DataLoader} in PyTorch \cite{pytorch} does not support assigning different numbers of mini-batches to CPUs and GPUs, or dynamically adjusting the batch size during GNN training;
thus, we further develop a \texttt{UnevenDDPIndices} wrapper to support the above features.

\begin{table*}[]
    \renewcommand{\arraystretch}{1}
    \caption{Epoch Time (sec) Comparison}
    \begin{tabularx}{\textwidth}{YYYY|YYY}
    \toprule
    \multicolumn{1}{l}{}        & \multicolumn{1}{c}{Sampler} & \multicolumn{1}{c}{Model}  & Protocol & Reddit             & Products               & MAG240M            \\ \midrule \midrule
    \multirow{7}{*}{Platform 1} & \multirow{4}{*}{Neighbor}   & \multirow{2}{*}{GCN}       & Standard & 3.27 (1.00$\times$)  & 5.20 (1.00$\times$) & 107.38 (1.00$\times$) \\
                                &                             &                            & Unified  & 2.32 (1.41$\times$)  & 3.95 (1.32$\times$) & 82.54 (1.30$\times$) \\
                                &                             & \multirow{2}{*}{GraphSAGE} & Standard & 3.11 (1.00$\times$)  & 4.94 (1.00$\times$) & 103.03 (1.00$\times$) \\
                                &                             &                            & Unified  & 2.41 (1.29$\times$)  & 3.85 (1.28$\times$) & 77.51 (1.33$\times$) \\
        {A100 (3g.20gb)}                        & \multirow{4}{*}{ShaDow}     & \multirow{2}{*}{GCN}       & Standard & 11.38 (1.00$\times$) & 41.50 (1.00$\times$) & 366.39 (1.00$\times$) \\
                                &                             &                            & Unified  & 9.32 (1.22$\times$)  & 34.30 (1.21$\times$) & 265.66 (1.38$\times$) \\
                                &                             & \multirow{2}{*}{GraphSAGE} & Standard & 10.59 (1.00$\times$) & 40.62 (1.00$\times$) & 445.80 (1.00$\times$) \\
                                &                             &                            & Unified  & 9.13 (1.16$\times$)  & 35.76 (1.14$\times$) & 350.91 (1.27$\times$) \\ \midrule
    \multirow{7}{*}{Platform 2} & \multirow{4}{*}{Neighbor}   & \multirow{2}{*}{GCN}       & Standard & 3.06 (1.00$\times$)  & 4.88 (1.00$\times$) & 100.32 (1.00$\times$)  \\
                                &                             &                            & Unified  & 2.43 (1.26$\times$)  & 4.08 (1.20$\times$) & 84.03 (1.19$\times$)  \\
                                &                             & \multirow{2}{*}{GraphSAGE} & Standard & 2.82 (1.00$\times$)  & 4.47 (1.00$\times$) & 93.12 (1.00$\times$)  \\
                                &                             &                            & Unified  & 2.37 (1.19$\times$)  & 3.78 (1.18$\times$) & 78.33 (1.19$\times$)  \\
        {A5000}                        & \multirow{4}{*}{ShaDow}     & \multirow{2}{*}{GCN}       & Standard & 8.96 (1.00$\times$)  & 34.88 (1.00$\times$) & 375.21 (1.00$\times$)  \\
                                &                             &                            & Unified  & 8.29 (1.08$\times$)  & 32.56 (1.07$\times$) & 306.80 (1.22$\times$)  \\
                                &                             & \multirow{2}{*}{GraphSAGE} & Standard & 9.71 (1.00$\times$)  & 37.34 (1.00$\times$) & 409.83 (1.00$\times$)  \\
                                &                             &                            & Unified  & 8.92 (1.09$\times$)  & 34.18 (1.09$\times$) & 337.07 (1.22$\times$) \\ \bottomrule
    \end{tabularx}
    \label{tab:overall_perf}
    \end{table*}

\subsection{Overall Performance}
In Table \ref{tab:overall_perf}, we compare the performance of our work, which adopts the Unified CPU-GPU training protocol and the optimizations mentioned in Section \ref{sec:opt}, with the DGL baseline, which adopts the Standard GNN training protocol (details in Figure \ref{fig:old_protocol}).
{The baseline designs are obtained from the official implementations of DGL \footnote{\url{https://github.com/dmlc/dgl/tree/master/examples/pytorch/ogb/ogbn-products}}\footnote{\url{https://github.com/dmlc/dgl/tree/master/examples/pytorch/ogb_lsc/MAG240M}}.
We then make minor code changes to the baseline designs to enable the Unified CPU-GPU training; the code changes are shown in Section \ref{sec:program_mod}.
We evaluate our protocol on two CPU-GPU platforms as described in Section \ref{sec:env}.
On Platform 1, our protocol leads to up to $1.41\times$ speedup compared with the Standard GNN training protocol that offloads most of the workload to the GPU.
On Platform 2, our protocol leads to up to $1.26\times$ speedup compared with the Standard GNN training protocol.}

\begin{figure}[t]
    \centering
    \includegraphics[width=8.6cm]{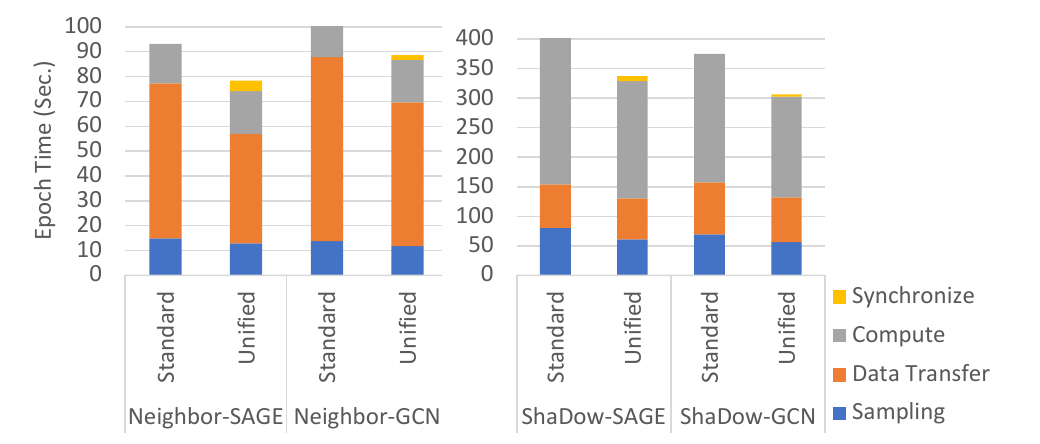}
    \caption{Epoch time breakdown}
    \vspace{-0.1cm}
     \label{fig:breakdown}
\end{figure}

{To understand the performance improvement of our protocol, we show the training time breakdown in Figure \ref{fig:breakdown} using MAG240M on Platform 2 as an example. 
For the Neighbor Sampling setup, our protocol leads to speedup primarily by reducing the CPU-GPU data transfer overhead.
For the ShaDow Sampling setup, our protocol achieves speedup mainly by reducing the workload on the GPU to reduce computation time.
The main difference between these two settings is that training with Neighbor Sampling is communication-intensive while training with ShaDow Sampling is computation-intensive.
This discrepancy arises from the density of subgraphs constructed by the two samplers:
The ShaDow Sampler constructs a localized subgraph for each node in a given mini-batch by inducing (i.e., sampling all the edges between a node set) from K-Hop neighboring nodes, whereas the Neighbor Sampler only samples edges along the sampling path of each mini-batch.
As a result, mini-batches constructed by the ShaDow Sampler hold significantly more edges than the mini-batches constructed by the Neighbor Sampler, which leads to higher computational complexity as the computation of GNNs is correlated with the number of edges in the mini-batch~\cite{dl_on_graphs}.
% Since the computation of most GNNs is correlated with the number of edges~\cite{dl_on_graphs}, training with Shadow Sampler leads to much higher computational complexity than with Neighbor Sampler. 
The experimental results show that our protocol can accelerate both computation-intensive and communication-intensive tasks, by reducing computation and data transfer time. respectively.}
{Additionally, as shown in Table \ref{util}, the Unified CPU-GPU protocol effectively improves resource utilization and memory bandwidth compared with the Standard protocol.
This is because our protocol exploits CPU compute cores to execute GNN operations, and also overlaps the computation with communications across multiple GNN training processes. 
To verify that our library works on platforms with multiple GPUs, we also experiment with 2 A5000 GPUs.
For such a platform where the GPUs are much faster than the CPUs, the speedup of the unified CPU-GPU protocol is limited (1.00$\times-1.07\times$).
This is because, after Dynamic Load Balancing, the CPU is only assigned a small portion of the workload, limiting the achievable speedup.
Similarly, using the Unified CPU-GPU training protocol on the A100 (no MIG enabled), the speedup is relatively limited compared with the Standard training protocol (1.04$\times-1.19\times$).}
{As mentioned in Section \ref{sec:intro}, our protocol is particularly useful in scenarios with constrained GPU resources; therefore, the limited speedup is expected when the GPU significantly outperforms the CPU.}

% Nevertheless, unified CPU-GPU training remains beneficial due to two key factors: (1) the increasing power of CPU platforms, and (2) the limited access to GPUs currently experienced by many users.

% Please add the following required packages to your document preamble:
% \usepackage{multirow}
\begin{table}[]
\caption{Utilization of CPU, GPU, and memory bandwidth}
\label{util}
\begin{tabular}{ccccc}
\hline
\multicolumn{1}{l}{Sampler-Model}          &   Protocol       & CPU     & GPU     & Mem. BW     \\ \hline \hline
\multirow{2}{*}{Neighbor-SAGE} & Standard & 2.00\%     & 25.50\% & 10.1 GB/sec \\ \cline{2-5} 
                               & Unified  & 25.10\% & 18.40\% & 21.2 GB/sec \\ \hline
\multirow{2}{*}{Neighbor-GCN}  & Standard & 1.70\%  & 28.40\% & 9.0 GB/sec    \\ \cline{2-5} 
                               & Unified  & 25.50\% & 24.10\% & 17.1 GB/sec \\ \hline
\multirow{2}{*}{ShaDow-SAGE}   & Standard & 2.54\%  & 37.37\% & 12.2 GB/sec \\ \cline{2-5} 
                               & Unified  & 24.80\% & 31.05\%    & 36.0 GB/sec   \\ \hline
\multirow{2}{*}{ShaDow-GCN}    & Standard & 2.34\%  & 35.30\% & 15.2 GB/sec   \\ \cline{2-5} 
                               & Unified  & 23.90\% & 31.70\% & 38.4 GB/sec   \\ \hline
\end{tabular}
\vspace{-0.1cm}
\end{table}

\begin{figure*}[t]
    \centering
    \includegraphics[width=16.2cm]{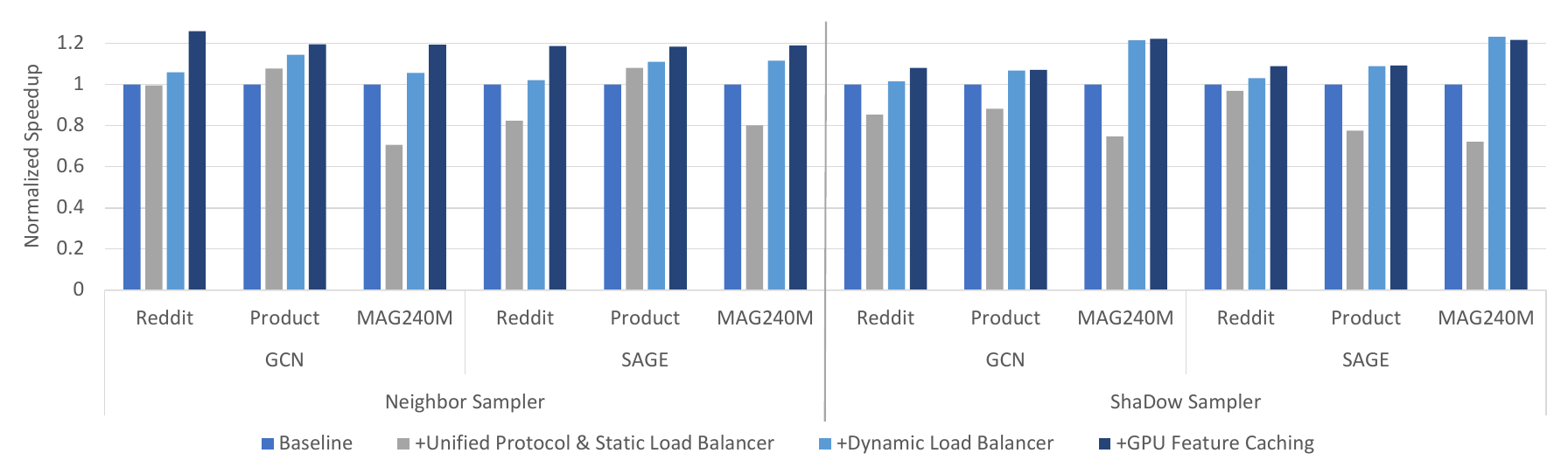}
    \caption{Impact of optimizations}
     \label{fig:ablation}
\end{figure*} 

% \subsection{Analysis of GNN Training Protocols}\label{sec:cpu_prof}
% To better understand the CPU implementation, we perform a detailed analysis using the PyTorch Profilier \cite{profiler} to output time traces of GNN training.
% We show the profiling results in Figure \ref{fig:breakdown}.

% We use the letter ``N" to refer to the Neighbor Sampler, and ``S" for the ShaDow Sampler; similarly, we use ``Pr" for the ogbn-products dataset, and ``Pa" for the ogbn-papers100M dataset.
% Using various samplers, models, and datasets, the profiling result shows that the memory allocation takes up $13\%-52\%$ of the training time.
% As explained in Section \ref{sec:malloc}, this is because the default serial memory allocator does not scale well for multi-threaded programs.
% Our library replaces the default memory allocator with TCMalloc \cite{tcmalloc}, a multi-threaded memory allocator, and achieves up to $1.8\times$ speedup compared with using the serial memory allocator.
% This is the first work that identifies memory allocation dominates a vast amount of time in GNN training on CPU platforms.

\subsection{Impact of Optimizations}\label{sec:ablation}
We evaluate the effectiveness of the optimizations.
{We start from the baseline design and gradually apply our optimizations to observe the performance improvements.
Figure \ref{fig:ablation} shows the speedup normalized with repect to the baseline design.
We first use the Unified CPU-GPU protocol, which uses the GNN Process Manager (see Section \ref{sec:multi_process}) to assign workload to both the CPU and GPU platform, and adjust workload assignment with a Static Load Balancer (see Section \ref{sec:manager}).
As mentioned in Section \ref{sec:manager}, static workload assignment only works if the workload of each mini-batch is similar, such as in the case of the ogbn-product dataset under the Neighbor Sampling algorithm.
For Reddit and MAG240M, the workload of each mini-batch varies and so a static assignment cannot balance the workload.
This causes the CPU platform to bottleneck the system, resulting in performance degradation.
After applying the Dynamic Load Balancer, which balances the workload based on the estimated workload for each mini-batch, our system is able to achieve performance improvement under various sampling algorithms, GNN models, and datasets.
Finally, we apply GPU feature caching, which reduces the overhead of accessing the unified shared memory, especially if there exist some frequently accessed nodes and edges in the graph, for example, the Reddit and MAG240M datasets under the Neighbor Sampling algorithm.}

\section{Related Work}
\subsection{Hybrid CPU-GPU Computation}
Several works have proposed to utilize both CPU and GPU to perform parallel computing.
CHC \cite{boostCUDA} exploits thread-level parallelism across the CPU and the GPU.
CHC achieves $1.42\times$ speedup on average compared with GPU-only baselines by utilizing the CPU cores to execute CUDA kernels.
FluidiCL \cite{FluidiCL} is a runtime system that takes an OpenCL program as input, and executes the program with both the CPU and the GPU.
SKMD \cite{skmd} proposes a single kernel multiple device framework that exploits data parallelism across the CPUs and the GPUs.
CHC, FluidCL, and SKMD are general frameworks that enable CPU-GPU computing for various applications.
However, in order to be generic, these frameworks only exploit parallelism within a single kernel to ensure the semantics of the executed program are preserved.
{As discussed in Section \ref{sec:ablation}, for GNN training, simply enabling CPU-GPU computing is not enough as it is non-trivial to achieve load balancing between the CPU and GPU.
Therefore, domain-specific optimizations like Dynamic Load Balancer is required to achieve performance improvment in terms of the overall training time.}

\subsection{GNN Training on Heterogeneous Platforms}
While most work on GNN training acceleration focus on using a homogeneous platform (e.g., GPU-only, CPU-only)~\cite{pyg,argo}, several works have exploited~\cite{graphact,dglv2,CARLA,hitgnn,hp-gnn} heterogeneous platforms such as CPU-GPU or CPU-FPGA platforms to accelerate GNN training.
GraphACT \cite{graphact} accelerates GNN training on a CPU-FPGA platform, and DistDGL\_v2 \cite{dglv2} accelerates GNN training on a CPU-GPU platform.
GraphACT and DistDGL\_v2 only assign lightweight operations such as mini-batch sampling to the CPU.
Thus, the CPU platform remains under-utilized.
HyScale-GNN \cite{hyscalegnn} accelerates GNN training on both CPU-GPU and CPU-FPGA platforms. 
It adopts the message-passing model for data communication among the devices, and the developer must explicitly handle numerous data transfers among the CPUs and GPUs or FPGAs during GNN training.
In contrast, our library takes advantage of the shared memory feature provided by state-of-the-art programming libraries, which reduces programming efforts by leaving the data transfer to the runtime. 
In addition, HyScale-GNN is neither compatible with existing GNN libraries nor open-sourced;
this limits its useability.

\section{Conclusion}
In this work, we proposed a Unified CPU-GPU training protocol for GNN training.
{The protocol explores the potential of incorporating CPU platform to improve GNN training performance.
Such an approach holds practical value in real-world applications, as CPU platforms are becoming increasingly powerful, while GPU resources often face limited availability due to high demand from many users or other researchers.}
By exploiting the computation resources and balancing the workload dynamically on both the CPUs and the GPUs, our work achieves up to $1.41\times$ speedup compared with the baseline design.
{While our protocol focused on CPU-GPU platforms in this work, it can be generalized to other accelerators.
For example, state-of-the-art FPGAs also support Unified Shared Memory (USM) for data communication  \cite{oneapi, vitis}, allowing it to be intergrated into the Unified training system.
With USM, Dynamic Load Balancing can be supported by collecting the GNN training time on the CPU and the FPGA to determine the workload ratio, and using the USM for mini-batch assignemnt.
In the future, we plan to generalize our work to support various types of accelerators, e.g., FPGA and TPU.}
% We also plan to extend our work to run on disaggregated datacenter \cite{diss_datacenter}, which allows users to dynamically reconfigure the hardware setup of the target platform.

% We discuss the caveats of our work.
% The first caveat is that our work adopts the synchronous SGD to train the GNN model, and assumes that the gradient calculation is independent between different mini-batch samples (see Section \ref{sec:algo});
% however, this assumption is false for some GNN sampling algorithms like GraphSAINT or Cluster-GCN \cite{clustergcn, graphsaint}.
% We plan to support these algorithms in the future by supporting data exchange among the sub-batches so that the samples do not need to be independent of each other.

% In addition, hybrid training splits the mini-batch into sub-batches, and distributes the sub-batches to CPUs and GPUs;
% this increases the total workload because there are fewer shared neighbors after the mini-batch is split into small sub-batches, which means fewer computed results can be reused.
% Thus, our library works best for the case when the GNN training performance on the CPU and the GPU is comparable.

\begin{acks}
This research was supported in part by a seed funding from Ershaghi Center for Energy Transitions, by the U.S. National Science Foundation (NSF) under grants CCF-1919289/SPX-2333009, CNS-2009057 and OAC-2209563, and by the DEVCOM Army Research Lab (ARL) under grant W911NF2220159.
\end{acks}

\bibliographystyle{ACM-Reference-Format}
\bibliography{ref}

%%% -*-BibTeX-*-
%%% Do NOT edit. File created by BibTeX with style
%%% ACM-Reference-Format-Journals [18-Jan-2012].

\begin{thebibliography}{45}

%%% ====================================================================
%%% NOTE TO THE USER: you can override these defaults by providing
%%% customized versions of any of these macros before the \bibliography
%%% command.  Each of them MUST provide its own final punctuation,
%%% except for \shownote{}, \showDOI{}, and \showURL{}.  The latter two
%%% do not use final punctuation, in order to avoid confusing it with
%%% the Web address.
%%%
%%% To suppress output of a particular field, define its macro to expand
%%% to an empty string, or better, \unskip, like this:
%%%
%%% \newcommand{\showDOI}[1]{\unskip}   % LaTeX syntax
%%%
%%% \def \showDOI #1{\unskip}           % plain TeX syntax
%%%
%%% ====================================================================

\ifx \showCODEN    \undefined \def \showCODEN     #1{\unskip}     \fi
\ifx \showDOI      \undefined \def \showDOI       #1{#1}\fi
\ifx \showISBNx    \undefined \def \showISBNx     #1{\unskip}     \fi
\ifx \showISBNxiii \undefined \def \showISBNxiii  #1{\unskip}     \fi
\ifx \showISSN     \undefined \def \showISSN      #1{\unskip}     \fi
\ifx \showLCCN     \undefined \def \showLCCN      #1{\unskip}     \fi
\ifx \shownote     \undefined \def \shownote      #1{#1}          \fi
\ifx \showarticletitle \undefined \def \showarticletitle #1{#1}   \fi
\ifx \showURL      \undefined \def \showURL       {\relax}        \fi
% The following commands are used for tagged output and should be
% invisible to TeX
\providecommand\bibfield[2]{#2}
\providecommand\bibinfo[2]{#2}
\providecommand\natexlab[1]{#1}
\providecommand\showeprint[2][]{arXiv:#2}

\bibitem[Chen et~al\mbox{.}(2016)]%
        {sgd}
\bibfield{author}{\bibinfo{person}{Jianmin Chen}, \bibinfo{person}{Rajat Monga}, \bibinfo{person}{Samy Bengio}, {and} \bibinfo{person}{Rafal Jozefowicz}.} \bibinfo{year}{2016}\natexlab{}.
\newblock \showarticletitle{Revisiting Distributed Synchronous SGD}. In \bibinfo{booktitle}{\emph{International Conference on Learning Representations Workshop}}.
\newblock


\bibitem[Chiang et~al\mbox{.}(2019)]%
        {clustergcn}
\bibfield{author}{\bibinfo{person}{Wei-Lin Chiang}, \bibinfo{person}{Xuanqing Liu}, \bibinfo{person}{Si Si}, \bibinfo{person}{Yang Li}, \bibinfo{person}{Samy Bengio}, {and} \bibinfo{person}{Cho-Jui Hsieh}.} \bibinfo{year}{2019}\natexlab{}.
\newblock \showarticletitle{Cluster-gcn: An efficient algorithm for training deep and large graph convolutional networks}. In \bibinfo{booktitle}{\emph{Proceedings of ACM SIGKDD International Conference on Knowledge Discovery \& Data Mining}}.
\newblock


\bibitem[Fey and Lenssen(2019)]%
        {pyg}
\bibfield{author}{\bibinfo{person}{Matthias Fey} {and} \bibinfo{person}{Jan~E. Lenssen}.} \bibinfo{year}{2019}\natexlab{}.
\newblock \showarticletitle{Fast Graph Representation Learning with {PyTorch Geometric}}. In \bibinfo{booktitle}{\emph{ICLR Workshop on Representation Learning on Graphs and Manifolds}}.
\newblock


\bibitem[Gandhi and Iyer(2021)]%
        {p3}
\bibfield{author}{\bibinfo{person}{Swapnil Gandhi} {and} \bibinfo{person}{Anand~Padmanabha Iyer}.} \bibinfo{year}{2021}\natexlab{}.
\newblock \showarticletitle{P3: Distributed Deep Graph Learning at Scale}. In \bibinfo{booktitle}{\emph{15th {USENIX} Symposium on Operating Systems Design and Implementation ({OSDI} 21)}}.
\newblock


\bibitem[Gilmer et~al\mbox{.}(2017)]%
        {message-passing}
\bibfield{author}{\bibinfo{person}{Justin Gilmer}, \bibinfo{person}{Samuel~S Schoenholz}, \bibinfo{person}{Patrick~F Riley}, \bibinfo{person}{Oriol Vinyals}, {and} \bibinfo{person}{George~E Dahl}.} \bibinfo{year}{2017}\natexlab{}.
\newblock \showarticletitle{Neural message passing for quantum chemistry}. In \bibinfo{booktitle}{\emph{International conference on machine learning}}. PMLR, \bibinfo{pages}{1263--1272}.
\newblock


\bibitem[Hamilton et~al\mbox{.}(2017)]%
        {graphsage}
\bibfield{author}{\bibinfo{person}{William~L Hamilton}, \bibinfo{person}{Rex Ying}, {and} \bibinfo{person}{Jure Leskovec}.} \bibinfo{year}{2017}\natexlab{}.
\newblock \showarticletitle{Inductive representation learning on large graphs}. In \bibinfo{booktitle}{\emph{Proceedings of the 31st International Conference on Neural Information Processing Systems}}.
\newblock


\bibitem[Hu et~al\mbox{.}(2021)]%
        {hu2021ogblsc}
\bibfield{author}{\bibinfo{person}{Weihua Hu}, \bibinfo{person}{Matthias Fey}, \bibinfo{person}{Hongyu Ren}, \bibinfo{person}{Maho Nakata}, \bibinfo{person}{Yuxiao Dong}, {and} \bibinfo{person}{Jure Leskovec}.} \bibinfo{year}{2021}\natexlab{}.
\newblock \showarticletitle{OGB-LSC: A Large-Scale Challenge for Machine Learning on Graphs}.
\newblock \bibinfo{journal}{\emph{arXiv preprint arXiv:2103.09430}} (\bibinfo{year}{2021}).
\newblock


\bibitem[Hu et~al\mbox{.}(2020)]%
        {ogb}
\bibfield{author}{\bibinfo{person}{Weihua Hu}, \bibinfo{person}{Matthias Fey}, \bibinfo{person}{Marinka Zitnik}, \bibinfo{person}{Yuxiao Dong}, \bibinfo{person}{Hongyu Ren}, \bibinfo{person}{Bowen Liu}, \bibinfo{person}{Michele Catasta}, {and} \bibinfo{person}{Jure Leskovec}.} \bibinfo{year}{2020}\natexlab{}.
\newblock \showarticletitle{Open Graph Benchmark: Datasets for Machine Learning on Graphs}.
\newblock \bibinfo{journal}{\emph{arXiv preprint arXiv:2005.00687}} (\bibinfo{year}{2020}).
\newblock


\bibitem[Kathail(2020)]%
        {vitis}
\bibfield{author}{\bibinfo{person}{Vinod Kathail}.} \bibinfo{year}{2020}\natexlab{}.
\newblock \showarticletitle{Xilinx vitis unified software platform}. In \bibinfo{booktitle}{\emph{Proceedings of the 2020 ACM/SIGDA International Symposium on Field-Programmable Gate Arrays}}. \bibinfo{pages}{173--174}.
\newblock


\bibitem[Kipf and Welling(2017)]%
        {gcn}
\bibfield{author}{\bibinfo{person}{Thomas~N. Kipf} {and} \bibinfo{person}{Max Welling}.} \bibinfo{year}{2017}\natexlab{}.
\newblock \showarticletitle{Semi-Supervised Classification with Graph Convolutional Networks}. In \bibinfo{booktitle}{\emph{International Conference on Learning Representations}}.
\newblock


\bibitem[Lee et~al\mbox{.}(2014)]%
        {boostCUDA}
\bibfield{author}{\bibinfo{person}{Changmin Lee}, \bibinfo{person}{Won~Woo Ro}, {and} \bibinfo{person}{Jean-Luc Gaudiot}.} \bibinfo{year}{2014}\natexlab{}.
\newblock \showarticletitle{Boosting CUDA Applications with CPU–GPU Hybrid Computing}. In \bibinfo{booktitle}{\emph{International Journal of Parallel Programming}}.
\newblock


\bibitem[Lee et~al\mbox{.}(2015)]%
        {skmd}
\bibfield{author}{\bibinfo{person}{Janghaeng Lee}, \bibinfo{person}{Mehrzad Samadi}, \bibinfo{person}{Yongjun Park}, {and} \bibinfo{person}{Scott Mahlke}.} \bibinfo{year}{2015}\natexlab{}.
\newblock \showarticletitle{SKMD: Single Kernel on Multiple Devices for Transparent CPU-GPU Collaboration}.
\newblock \bibinfo{journal}{\emph{ACM Trans. Comput. Syst.}} (\bibinfo{year}{2015}).
\newblock


\bibitem[Li et~al\mbox{.}(2020a)]%
        {nvlink}
\bibfield{author}{\bibinfo{person}{Ang Li}, \bibinfo{person}{Shuaiwen~Leon Song}, \bibinfo{person}{Jieyang Chen}, \bibinfo{person}{Jiajia Li}, \bibinfo{person}{Xu Liu}, \bibinfo{person}{Nathan~R. Tallent}, {and} \bibinfo{person}{Kevin~J. Barker}.} \bibinfo{year}{2020}\natexlab{a}.
\newblock \showarticletitle{Evaluating Modern GPU Interconnect: PCIe, NVLink, NV-SLI, NVSwitch and GPUDirect}.
\newblock \bibinfo{journal}{\emph{IEEE Transactions on Parallel and Distributed Systems (TPDS)}} (\bibinfo{year}{2020}).
\newblock


\bibitem[Li et~al\mbox{.}(2020b)]%
        {ddp}
\bibfield{author}{\bibinfo{person}{Shen Li}, \bibinfo{person}{Yanli Zhao}, \bibinfo{person}{Rohan Varma}, \bibinfo{person}{Omkar Salpekar}, \bibinfo{person}{Pieter Noordhuis}, \bibinfo{person}{Teng Li}, \bibinfo{person}{Adam Paszke}, \bibinfo{person}{Jeff Smith}, \bibinfo{person}{Brian Vaughan}, \bibinfo{person}{Pritam Damania}, {and} \bibinfo{person}{Soumith Chintala}.} \bibinfo{year}{2020}\natexlab{b}.
\newblock \showarticletitle{PyTorch Distributed: Experiences on Accelerating Data Parallel Training}.
\newblock \bibinfo{journal}{\emph{Proceedings of the VLDB Endowment}} (\bibinfo{year}{2020}).
\newblock


\bibitem[Lin et~al\mbox{.}(2024)]%
        {argo}
\bibfield{author}{\bibinfo{person}{Yi-Chien Lin}, \bibinfo{person}{Yuyang Chen}, \bibinfo{person}{Sameh Gobriel}, \bibinfo{person}{Nilesh Jain}, \bibinfo{person}{Gopi~Krishna Jha}, {and} \bibinfo{person}{Viktor Prasanna}.} \bibinfo{year}{2024}\natexlab{}.
\newblock \showarticletitle{ARGO: An Auto-Tuning Runtime System for Scalable GNN Training on Multi-Core Processor}.
\newblock
\showeprint[arxiv]{2402.03671}~[cs.DC]


\bibitem[Lin and Prasanna(2023)]%
        {hyscalegnn}
\bibfield{author}{\bibinfo{person}{Yi-Chien Lin} {and} \bibinfo{person}{Viktor Prasanna}.} \bibinfo{year}{2023}\natexlab{}.
\newblock \showarticletitle{HyScale-GNN: A Scalable Hybrid GNN Training System on Single-Node Heterogeneous Architecture}. In \bibinfo{booktitle}{\emph{International Parallel and Distributed Processing Symposium}}.
\newblock


\bibitem[Lin et~al\mbox{.}(2022b)]%
        {CARLA}
\bibfield{author}{\bibinfo{person}{Yi-Chien Lin}, \bibinfo{person}{Bingyi Zhang}, {and} \bibinfo{person}{Viktor Prasanna}.} \bibinfo{year}{2022}\natexlab{b}.
\newblock \showarticletitle{Accelerating GNN Training on CPU+Multi-FPGA Heterogeneous Platform}. In \bibinfo{booktitle}{\emph{High Performance Computing}}. \bibinfo{publisher}{Springer International Publishing}.
\newblock


\bibitem[Lin et~al\mbox{.}(2022c)]%
        {hp-gnn}
\bibfield{author}{\bibinfo{person}{Yi-Chien Lin}, \bibinfo{person}{Bingyi Zhang}, {and} \bibinfo{person}{Viktor Prasanna}.} \bibinfo{year}{2022}\natexlab{c}.
\newblock \showarticletitle{HP-GNN: Generating High Throughput GNN Training Implementation on CPU-FPGA Heterogeneous Platform}. In \bibinfo{booktitle}{\emph{International Symposium on Field-Programmable Gate Arrays}}.
\newblock


\bibitem[Lin et~al\mbox{.}(2023)]%
        {hitgnn}
\bibfield{author}{\bibinfo{person}{Yi-Chien Lin}, \bibinfo{person}{Bingyi Zhang}, {and} \bibinfo{person}{Viktor Prasanna}.} \bibinfo{year}{2023}\natexlab{}.
\newblock \showarticletitle{HitGNN: High-throughput GNN Training Framework on CPU+Multi-FPGA Heterogeneous Platform}.
\newblock
\showeprint[arxiv]{2303.01568}~[cs.DC]


\bibitem[Lin et~al\mbox{.}(2022a)]%
        {DLRM_GPU}
\bibfield{author}{\bibinfo{person}{Zhongyi Lin}, \bibinfo{person}{Louis Feng}, \bibinfo{person}{Ehsan~K. Ardestani}, \bibinfo{person}{Jaewon Lee}, \bibinfo{person}{John Lundell}, \bibinfo{person}{Changkyu Kim}, \bibinfo{person}{Arun Kejariwal}, {and} \bibinfo{person}{John~D. Owens}.} \bibinfo{year}{2022}\natexlab{a}.
\newblock \showarticletitle{Building a Performance Model for Deep Learning Recommendation Model Training on GPUs}. In \bibinfo{booktitle}{\emph{2022 IEEE International Symposium on Performance Analysis of Systems and Software (ISPASS)}}.
\newblock


\bibitem[Lin et~al\mbox{.}(2020)]%
        {pagraph}
\bibfield{author}{\bibinfo{person}{Zhiqi Lin}, \bibinfo{person}{Cheng Li}, \bibinfo{person}{Youshan Miao}, \bibinfo{person}{Yunxin Liu}, {and} \bibinfo{person}{Yinlong Xu}.} \bibinfo{year}{2020}\natexlab{}.
\newblock \showarticletitle{PaGraph: Scaling GNN Training on Large Graphs via Computation-Aware Caching}. In \bibinfo{booktitle}{\emph{Proceedings of the 11th ACM Symposium on Cloud Computing}}.
\newblock


\bibitem[Lopera et~al\mbox{.}(2021)]%
        {gnn-eda}
\bibfield{author}{\bibinfo{person}{Daniela~S{\'a}nchez Lopera}, \bibinfo{person}{Lorenzo Servadei}, \bibinfo{person}{Gamze~Naz Kiprit}, \bibinfo{person}{Souvik Hazra}, \bibinfo{person}{Robert Wille}, {and} \bibinfo{person}{Wolfgang Ecker}.} \bibinfo{year}{2021}\natexlab{}.
\newblock \showarticletitle{A survey of graph neural networks for electronic design automation}. In \bibinfo{booktitle}{\emph{Workshop on Machine Learning for CAD (MLCAD)}}. IEEE.
\newblock


\bibitem[Md et~al\mbox{.}(2021)]%
        {distgnn}
\bibfield{author}{\bibinfo{person}{Vasimuddin Md}, \bibinfo{person}{Sanchit Misra}, \bibinfo{person}{Guixiang Ma}, \bibinfo{person}{Ramanarayan Mohanty}, \bibinfo{person}{Evangelos Georganas}, \bibinfo{person}{Alexander Heinecke}, \bibinfo{person}{Dhiraj~D. Kalamkar}, \bibinfo{person}{Nesreen~K. Ahmed}, {and} \bibinfo{person}{Sasikanth Avancha}.} \bibinfo{year}{2021}\natexlab{}.
\newblock \showarticletitle{DistGNN: Scalable Distributed Training for Large-Scale Graph Neural Networks}. In \bibinfo{booktitle}{\emph{Proceedings of the International Conference for High Performance Computing, Networking, Storage and Analysis (SC)}}.
\newblock


\bibitem[Nozal and Bosque(2021)]%
        {oneapi}
\bibfield{author}{\bibinfo{person}{Ra{\'u}l Nozal} {and} \bibinfo{person}{Jose~Luis Bosque}.} \bibinfo{year}{2021}\natexlab{}.
\newblock \showarticletitle{Exploiting Co-execution with OneAPI: Heterogeneity from a Modern Perspective}. In \bibinfo{booktitle}{\emph{Euro-Par 2021: Parallel Processing}}.
\newblock


\bibitem[NVIDIA({[n.\,d.]})]%
        {nvsmi}
\bibfield{author}{\bibinfo{person}{NVIDIA}.} \bibinfo{year}{[n.\,d.]}\natexlab{}.
\newblock \bibinfo{booktitle}{\emph{System Management Interface [Online]}}.
\newblock
\urldef\tempurl%
\url{{https://developer.nvidia.com/nvidia-system-management-interface}}
\showURL{%
\tempurl}
\newblock
\shownote{Accessed: 2023-09-05}.


\bibitem[Pandit and Govindarajan(2014)]%
        {FluidiCL}
\bibfield{author}{\bibinfo{person}{Prasanna Pandit} {and} \bibinfo{person}{R. Govindarajan}.} \bibinfo{year}{2014}\natexlab{}.
\newblock \showarticletitle{Fluidic Kernels: Cooperative Execution of OpenCL Programs on Multiple Heterogeneous Devices}. In \bibinfo{booktitle}{\emph{Proceedings of Annual IEEE/ACM International Symposium on Code Generation and Optimization}} (Orlando, FL, USA) \emph{(\bibinfo{series}{CGO '14})}. \bibinfo{pages}{273–283}.
\newblock


\bibitem[Paszke et~al\mbox{.}(2019)]%
        {pytorch}
\bibfield{author}{\bibinfo{person}{Adam Paszke}, \bibinfo{person}{Sam Gross}, \bibinfo{person}{Francisco Massa}, \bibinfo{person}{Adam Lerer}, \bibinfo{person}{James Bradbury}, \bibinfo{person}{Gregory Chanan}, \bibinfo{person}{Trevor Killeen}, \bibinfo{person}{Zeming Lin}, \bibinfo{person}{Natalia Gimelshein}, \bibinfo{person}{Luca Antiga}, {et~al\mbox{.}}} \bibinfo{year}{2019}\natexlab{}.
\newblock \showarticletitle{Pytorch: An imperative style, high-performance deep learning library}.
\newblock \bibinfo{journal}{\emph{Advances in neural information processing systems}}  \bibinfo{volume}{32} (\bibinfo{year}{2019}).
\newblock


\bibitem[Reinders(2005)]%
        {vtune}
\bibfield{author}{\bibinfo{person}{James Reinders}.} \bibinfo{year}{2005}\natexlab{}.
\newblock \bibinfo{booktitle}{\emph{VTune performance analyzer essentials}}. Vol.~\bibinfo{volume}{9}.
\newblock \bibinfo{publisher}{Intel Press Santa Clara}.
\newblock


\bibitem[Tan et~al\mbox{.}(2021)]%
        {MIG}
\bibfield{author}{\bibinfo{person}{Cheng Tan}, \bibinfo{person}{Zhichao Li}, \bibinfo{person}{Jian Zhang}, \bibinfo{person}{Yu Cao}, \bibinfo{person}{Sikai Qi}, \bibinfo{person}{Zherui Liu}, \bibinfo{person}{Yibo Zhu}, {and} \bibinfo{person}{Chuanxiong Guo}.} \bibinfo{year}{2021}\natexlab{}.
\newblock \showarticletitle{Serving DNN models with multi-instance gpus: A case of the reconfigurable machine scheduling problem}.
\newblock \bibinfo{journal}{\emph{arXiv preprint arXiv:2109.11067}} (\bibinfo{year}{2021}).
\newblock


\bibitem[Thorpe et~al\mbox{.}(2021)]%
        {dorylus}
\bibfield{author}{\bibinfo{person}{John Thorpe}, \bibinfo{person}{Yifan Qiao}, \bibinfo{person}{Jonathan Eyolfson}, \bibinfo{person}{Shen Teng}, \bibinfo{person}{Guanzhou Hu}, \bibinfo{person}{Zhihao Jia}, \bibinfo{person}{Jinliang Wei}, \bibinfo{person}{Keval Vora}, \bibinfo{person}{Ravi Netravali}, \bibinfo{person}{Miryung Kim}, {and} \bibinfo{person}{Guoqing~Harry Xu}.} \bibinfo{year}{2021}\natexlab{}.
\newblock \showarticletitle{Dorylus: Affordable, scalable, and accurate GNN training with distributed CPU servers and serverless threads}. In \bibinfo{booktitle}{\emph{15th USENIX Symposium on Operating Systems Design and Implementation (OSDI 21)}}.
\newblock


\bibitem[Veličković et~al\mbox{.}(2018)]%
        {gat}
\bibfield{author}{\bibinfo{person}{Petar Veličković}, \bibinfo{person}{Guillem Cucurull}, \bibinfo{person}{Arantxa Casanova}, \bibinfo{person}{Adriana Romero}, \bibinfo{person}{Pietro Liò}, {and} \bibinfo{person}{Yoshua Bengio}.} \bibinfo{year}{2018}\natexlab{}.
\newblock \showarticletitle{Graph Attention Networks}. In \bibinfo{booktitle}{\emph{International Conference on Learning Representations}}.
\newblock
\urldef\tempurl%
\url{https://openreview.net/forum?id=rJXMpikCZ}
\showURL{%
\tempurl}


\bibitem[Wang et~al\mbox{.}(2019)]%
        {dgl}
\bibfield{author}{\bibinfo{person}{Minjie Wang}, \bibinfo{person}{Da Zheng}, \bibinfo{person}{Zihao Ye}, \bibinfo{person}{Quan Gan}, \bibinfo{person}{Mufei Li}, \bibinfo{person}{Xiang Song}, \bibinfo{person}{Jinjing Zhou}, \bibinfo{person}{Chao Ma}, \bibinfo{person}{Lingfan Yu}, \bibinfo{person}{Yu Gai}, \bibinfo{person}{Tianjun Xiao}, \bibinfo{person}{Tong He}, \bibinfo{person}{George Karypis}, \bibinfo{person}{Jinyang Li}, {and} \bibinfo{person}{Zheng Zhang}.} \bibinfo{year}{2019}\natexlab{}.
\newblock \showarticletitle{Deep Graph Library: A Graph-Centric, Highly-Performant Package for Graph Neural Networks}.
\newblock \bibinfo{journal}{\emph{arXiv preprint arXiv:1909.01315}} (\bibinfo{year}{2019}).
\newblock


\bibitem[Wang et~al\mbox{.}(2022)]%
        {hugeCTR}
\bibfield{author}{\bibinfo{person}{Zehuan Wang}, \bibinfo{person}{Yingcan Wei}, \bibinfo{person}{Minseok Lee}, \bibinfo{person}{Matthias Langer}, \bibinfo{person}{Fan Yu}, \bibinfo{person}{Jie Liu}, \bibinfo{person}{Shijie Liu}, \bibinfo{person}{Daniel~G. Abel}, \bibinfo{person}{Xu Guo}, \bibinfo{person}{Jianbing Dong}, \bibinfo{person}{Ji Shi}, {and} \bibinfo{person}{Kunlun Li}.} \bibinfo{year}{2022}\natexlab{}.
\newblock \showarticletitle{Merlin HugeCTR: GPU-Accelerated Recommender System Training and Inference}. In \bibinfo{booktitle}{\emph{Proceedings of the 16th ACM Conference on Recommender Systems}} (Seattle, WA, USA) \emph{(\bibinfo{series}{RecSys '22})}. \bibinfo{publisher}{Association for Computing Machinery}, \bibinfo{address}{New York, NY, USA}, \bibinfo{pages}{534–537}.
\newblock
\showISBNx{9781450392785}
\urldef\tempurl%
\url{https://doi.org/10.1145/3523227.3547405}
\showDOI{\tempurl}


\bibitem[Wu et~al\mbox{.}(2022)]%
        {eda2}
\bibfield{author}{\bibinfo{person}{Nan Wu}, \bibinfo{person}{Hang Yang}, \bibinfo{person}{Yuan Xie}, \bibinfo{person}{Pan Li}, {and} \bibinfo{person}{Cong Hao}.} \bibinfo{year}{2022}\natexlab{}.
\newblock \showarticletitle{High-Level Synthesis Performance Prediction Using GNNs: Benchmarking, Modeling, and Advancing}. In \bibinfo{booktitle}{\emph{Proceedings of the Design Automation Conference}}.
\newblock


\bibitem[Xu et~al\mbox{.}(2019)]%
        {gin}
\bibfield{author}{\bibinfo{person}{Keyulu Xu}, \bibinfo{person}{Weihua Hu}, \bibinfo{person}{Jure Leskovec}, {and} \bibinfo{person}{Stefanie Jegelka}.} \bibinfo{year}{2019}\natexlab{}.
\newblock \showarticletitle{How Powerful are Graph Neural Networks?}. In \bibinfo{booktitle}{\emph{International Conference on Learning Representations}}.
\newblock
\urldef\tempurl%
\url{https://openreview.net/forum?id=ryGs6iA5Km}
\showURL{%
\tempurl}


\bibitem[Yang and Li(2023)]%
        {yang_li_2023}
\bibfield{author}{\bibinfo{person}{Chu-I Yang} {and} \bibinfo{person}{Yi-Pei Li}.} \bibinfo{year}{2023}\natexlab{}.
\newblock \showarticletitle{Explainable uncertainty quantifications for deep learning-based molecular property prediction}.
\newblock \bibinfo{journal}{\emph{Journal of Cheminformatics}} (\bibinfo{year}{2023}).
\newblock


\bibitem[Yang et~al\mbox{.}(2022)]%
        {gnnlab}
\bibfield{author}{\bibinfo{person}{Jianbang Yang}, \bibinfo{person}{Dahai Tang}, \bibinfo{person}{Xiaoniu Song}, \bibinfo{person}{Lei Wang}, \bibinfo{person}{Qiang Yin}, \bibinfo{person}{Rong Chen}, \bibinfo{person}{Wenyuan Yu}, {and} \bibinfo{person}{Jingren Zhou}.} \bibinfo{year}{2022}\natexlab{}.
\newblock \showarticletitle{GNNLab: A Factored System for Sample-Based GNN Training over GPUs}. In \bibinfo{booktitle}{\emph{Proceedings of the 17th European Conference on Computer Systems}}.
\newblock


\bibitem[Ying et~al\mbox{.}(2018)]%
        {recommend1}
\bibfield{author}{\bibinfo{person}{Rex Ying}, \bibinfo{person}{Ruining He}, \bibinfo{person}{Kaifeng Chen}, \bibinfo{person}{Pong Eksombatchai}, \bibinfo{person}{William~L Hamilton}, {and} \bibinfo{person}{Jure Leskovec}.} \bibinfo{year}{2018}\natexlab{}.
\newblock \showarticletitle{Graph convolutional neural networks for web-scale recommender systems}. In \bibinfo{booktitle}{\emph{Proceedings of the 24th ACM International Conference on Knowledge Discovery \& Data Mining}}.
\newblock


\bibitem[Zeng and Prasanna(2020)]%
        {graphact}
\bibfield{author}{\bibinfo{person}{Hanqing Zeng} {and} \bibinfo{person}{Viktor Prasanna}.} \bibinfo{year}{2020}\natexlab{}.
\newblock \showarticletitle{GraphACT: Accelerating GCN training on CPU-FPGA heterogeneous platforms}. In \bibinfo{booktitle}{\emph{Proceedings of the 2020 ACM/SIGDA International Symposium on Field-Programmable Gate Arrays}}.
\newblock


\bibitem[Zeng et~al\mbox{.}(2021)]%
        {shaDow}
\bibfield{author}{\bibinfo{person}{Hanqing Zeng}, \bibinfo{person}{Muhan Zhang}, \bibinfo{person}{Yinglong Xia}, \bibinfo{person}{Ajitesh Srivastava}, \bibinfo{person}{Andrey Malevich}, \bibinfo{person}{Rajgopal Kannan}, \bibinfo{person}{Viktor Prasanna}, \bibinfo{person}{Long Jin}, {and} \bibinfo{person}{Ren Chen}.} \bibinfo{year}{2021}\natexlab{}.
\newblock \showarticletitle{Decoupling the Depth and Scope of Graph Neural Networks}. In \bibinfo{booktitle}{\emph{Advances in Neural Information Processing Systems}}.
\newblock


\bibitem[Zeng et~al\mbox{.}(2020)]%
        {graphsaint}
\bibfield{author}{\bibinfo{person}{Hanqing Zeng}, \bibinfo{person}{Hongkuan Zhou}, \bibinfo{person}{Ajitesh Srivastava}, \bibinfo{person}{Rajgopal Kannan}, {and} \bibinfo{person}{Viktor Prasanna}.} \bibinfo{year}{2020}\natexlab{}.
\newblock \showarticletitle{{GraphSAINT}: Graph Sampling Based Inductive Learning Method}. In \bibinfo{booktitle}{\emph{International Conference on Learning Representations}}.
\newblock


\bibitem[Zhang et~al\mbox{.}(2022)]%
        {dl_on_graphs}
\bibfield{author}{\bibinfo{person}{Z. Zhang}, \bibinfo{person}{P. Cui}, {and} \bibinfo{person}{W. Zhu}.} \bibinfo{year}{2022}\natexlab{}.
\newblock \showarticletitle{Deep Learning on Graphs: A Survey}.
\newblock \bibinfo{journal}{\emph{IEEE Transactions on Knowledge \& Data Engineering}} (\bibinfo{date}{jan} \bibinfo{year}{2022}).
\newblock
\urldef\tempurl%
\url{https://doi.org/10.1109/TKDE.2020.2981333}
\showDOI{\tempurl}


\bibitem[Zheng et~al\mbox{.}(2020)]%
        {distdgl}
\bibfield{author}{\bibinfo{person}{Da Zheng} {et~al\mbox{.}}} \bibinfo{year}{2020}\natexlab{}.
\newblock \showarticletitle{Distdgl: distributed graph neural network training for billion-scale graphs}. In \bibinfo{booktitle}{\emph{2020 IEEE/ACM 10th Workshop on Irregular Applications: Architectures and Algorithms (IA3)}}.
\newblock


\bibitem[Zheng et~al\mbox{.}(2022)]%
        {dglv2}
\bibfield{author}{\bibinfo{person}{Da Zheng} {et~al\mbox{.}}} \bibinfo{year}{2022}\natexlab{}.
\newblock \showarticletitle{Distributed Hybrid CPU and GPU Training for Graph Neural Networks on Billion-Scale Heterogeneous Graphs}. In \bibinfo{booktitle}{\emph{ACM SIGKDD Conference on Knowledge Discovery and Data Mining}}.
\newblock


\bibitem[Zhu et~al\mbox{.}(2019)]%
        {recommend2}
\bibfield{author}{\bibinfo{person}{Rong Zhu}, \bibinfo{person}{Kun Zhao}, \bibinfo{person}{Hongxia Yang}, \bibinfo{person}{Wei Lin}, \bibinfo{person}{Chang Zhou}, \bibinfo{person}{Baole Ai}, \bibinfo{person}{Yong Li}, {and} \bibinfo{person}{Jingren Zhou}.} \bibinfo{year}{2019}\natexlab{}.
\newblock \showarticletitle{AliGraph: a comprehensive graph neural network platform}.
\newblock \bibinfo{journal}{\emph{Proceedings of the VLDB Endowment}} (\bibinfo{year}{2019}).
\newblock


\end{thebibliography}

\end{document}